\documentclass[12pt,fleqn]{article}

\usepackage[dvips]{graphics}
\newcommand {\be}{\begin{equation}}
\newcommand {\ee}{\end{equation}}
\newcommand {\beqn}{\begin{eqnarray}}
\newcommand {\eeqn}{\end{eqnarray}}
\newcommand {\f}{\frac}
\newcommand {\nn}{\nonumber}
\newcommand {\p}{\partial}

\newcommand\TL{\hfil$\displaystyle{##}$}
\newcommand\TR{$\displaystyle{{}##}$\hfil}
\newcommand\TC{\hfil$\displaystyle{##}$\hfil}

\def\seqalign#1#2{\vcenter{\openup1\jot
  \halign{\strut #1\cr #2 \cr}}}
\def\lbldef#1#2{\expandafter\gdef\csname #1\endcsname {#2}}
\newcommand{\eqn}[3][]{\lbldef{#2}{(\ref{#2})}%
\def\@eqnstyle{#1}%
\ifx\@eqnstyle\@empty%
\begin{equation} \eqalign{#3} \label{#2} \end{equation}%
\else%
\begin{equation} \seqalign{\span\TC}{#3} \label{#2} \end{equation}%
\fi}
\def\eqalign#1{\vcenter{\openup1\jot
    \halign{\strut\span\TL & \span\TR\cr #1 \cr
   }}}

\def\p2{{p \over 2}}

\input epsf

\begin{document}


\thispagestyle{empty}
{\hfill  \parbox{8cm} {WIS/13/06-SEPT-DPP$~$hep-th/0608229 }}\\

\begin{center} \noindent \Large \bf
  Chiral symmetry breaking and restoration from holography
\end{center}

\vskip 0.1cm
\begin{center}
{ \normalsize \bf D. Gepner and Shesansu Sekahr Pal}\\
{ Weizmann institute of  science, 76100 Rehovot, Israel \\
\sf doron$\cdot$gepner ${\frame{\shortstack{AT}}}$~~~~ weizmann$\cdot$ac$\cdot$il\\
\sf shesansu$\cdot$pal ${\frame{\shortstack{AT}}}$~~~~ weizmann$\cdot$ac$\cdot$il}
\end{center}

\centerline{\bf \small Abstract}
We study  intersection of $N_c$  color D4 branes with topology 
$R^{1,3}\times S^1$ with  radius $R_{\tau}$  of $S^1$ 
with $N_f$ Dp-branes and
 anti-Dp branes  in the strong coupling limit in the probe approximation. 
The resulting model has $U(N_f)\times U(N_f)$ global symmetry.    
This is interpreted as the chiral symmetry group in this geometric setting. 
We see an $n$ dimensional
theory for $n$ overlapping directions between color and flavor branes. 
At zero temperature we do
see the breakdown of chiral symmetry, but there arises a puzzle: we do not 
see any massless NG boson to the break down of this 
symmetry group for $n=2,3$ for a specific p. At finite temperature we do see 
the restoration 
of chiral  symmetry group and the associated phase transitions are of first 
order. The asymptotic distance of separation, L,   between the quarks 
increases as one moves deep into the bulk at zero temperature case. 
At finite temperature the distance between the quarks do not vary much 
but  as one comes
close to the horizon the distance starts to decrease very fast.
The chiral symmetry restoration is described by a curve, which connects
L with the 
temperature, T. In general this quantity is very difficult to compute but if
we evaluate it numerically then the curve is described by an equation $L T=c$,
where c is a constant and is much smaller than one. It means  for $L/R_{\tau}$ 
above $2\pi c$ there occurs the deconfined phase along with the chiral symmetry
restored phase. \\
We also study another system in which we have added an electric field 
to the color D4-branes at finite temperature. This model shows up chiral 
symmetry breaking depending on the value of the electric field and a 
dimensionless parameter, which is related to the temperature of the background. 
But in this case there appears a puzzle: it exhibits chiral symmetry 
restoration in the confining phase.

\newpage

\section{Introduction}
The study of ``QCD'' through ADS/CFT has been an interesting step to 
understand some of the mysteries of the universe. This QCD that we are after
do not looks to be in the same class of real QCD. From the study of strong 
nuclear interaction there are three important properties that we need to 
understand better. Those are 
Confinement, Chiral symmetry breaking  and asymptotic freedom and most
importantly all these in the limit of small number of color degrees of freedom.
However, our study till now has been limited only to the case when we have 
large number color degrees of freedom i.e. in the  class of Maldacena limit
or duality. In this latter class of models where we try to view all these 
interesting properties obeyed by the strong nuclear interaction in terms of 
geometric language and this geometric language is necessarily a low energy 
effective language. So it means that it is harder to see asymptotic freedom 
in this language. More importantly, to reproduce real QCD means one need to 
go to small  
 color limit. Nevertheless the  confinement and chiral symmetry 
breaking can be studied at low energy effective theories  
 in the large color limit. Even with this shortcomings of large color 
limit where we have to sacrifice one of the interesting properties of strong
nuclear interaction in the geometric sense, we still have the other two 
interesting properties with us and the hope is that 
one might be able to understand asymptotic freedom of QCD by making an ultraviolet completion of this theory i.e. by embedding this low energy 
effective theory
described in a geometrical way in a dual theory which has got a nice UV 
behavior 
in the sense of seeing the asymptotic freedom. If that is going to be case then
why not try to understand in a much better way the properties that one would 
like to see in the low energy effective theory.  
With this aim to understand the confinement and chiral symmetry breaking let us 
proceed and  get some understanding of these phenomena and their phase diagram by considering some toy models \cite{ew,ss}. Before getting into that we 
should not forget 
that we are studying these two phenomena in a theory that is plagued by the 
presence of Kaluza-Klein modes.
In order to get a pure gluonic theory one need a procedure to  make  
these KK modes   heavy and can be integrated out, but in practice its becoming 
very difficult to achieve \cite{gn, ssp, lm1}, which is 
studied in the context of \cite{lm}. Hence, we shall assume 
that they are not there or
are integrated out in some region of parameter space.
 With all these drawbacks in mind we hope to understand some of mass formula, 
relation among couplings and some other properties that could be useful in 
trying to understand the real QCD. 

The model that we shall be considering are those that are described in a 
geometrical way in the presence of fluxes coming from either NS-NS or 
R-R sector of closed string or both. So, the degrees of freedom are those
coming from metric, $g_{MN}$, dilaton, and various p-from field strength. We
shall call these as the background geometry which may or may not be showing 
confinement. This background geometry is obtained by taking the near horizon 
limit of $N_c$ Dp-branes in the  large 't Hooft coupling limit and 
large  $N_c$ limit.
For this background geometry the fields that reside on the world volume of
the p-brane transforms adjointly 
under the color group.

To add matter fields which transform (bi-)fundamentally a prescription was 
proposed by  Karch-Katz \cite{kk}. This involves
adding  some branes to this background geometry. The result of this is that 
the fields that reside on the open string that stretch between these two 
branes transform under the product gauge group as bifundamentally. The 
addition 
of branes is going to change the energy momentum of the system and will 
back react on the background. In order to avoid it we shall take the number 
of these later  branes to be very small in comparison to the branes 
that gives us
the background geometry. This approximation is called as probe-brane 
approximation \cite{beegk, ts}-\cite{ss}.
In this setting the chiral symmetry  breaking/restoration is 
understood in two different ways for a specific background geometry. In one 
case \cite{kmmw1} the common directions along which both the color and 
probe (flavor)
 branes are not extended correspond to the plane of rotation of some group
and it is this global rotation group correspond to the chiral symmetry group. 
For an example, this group is U(1) if the background geometry is described by
the near horizon limit of a stack D4 branes and the flavor brane is a bunch
of D6 branes with 4 overlapping directions and 4 non-overlapping directions.  
In the other case \cite{ss} the chiral symmetry group is again a global 
symmetry group
but realized as the gauge symmetry of the flavor branes. In this paper we 
shall use 
the later approach to chiral symmetry group to study some phase 
diagram involving chiral
symmetry breaking/restoration  and confinement/deconfinement transition.
To understand chiral symmetry breaking/restoration, using this 
geometric language, we need to compute the action of the flavor brane 
with different embeddings 
which will give us information about this symmetry. 

By taking the background geometry as that of the near horizon limit of D4 
branes \cite{ew} gives us a confining theory which is geometrically 
described in terms
of metric, dilaton and a 4-form magnetic flux. This smooth background has an 
interesting property which is not shared by any other known confining 
background geometry is that the metric is completely diagonal and computation 
in this geometry becomes very easy.  
The radial coordinate do not stay in the range from zero to
infinity rather from some non-zero value $u_{0}$ to $\infty$. This is taken 
to avoid closed time like curves in the background. The topology of this 
geometry roughly is $R^{1,3}\times S^1\times R\times S^4 $, where the color D4 branes are extended along $R^{1,3}$ and wrapped around the $S^1$. Now putting 
a bunch of probe Dp-branes whose worldvolume directions can be extended 
along any directions except the $S^1$, which we have mentioned very clearly 
later. 
The study of this probe brane is achieved by taking a 
bunch of Dp-${\bar Dp}$ branes and it  gives an interesting way to break the 
chiral symmetry. 
This symmetry  is generated by realizing the gauge symmetry of $N_f$ Dp and 
${\bar Dp}$ branes, which is $U(N_f)\times U(N_f)$. The break down of chiral 
symmetry means breaking this product group down to its diagonal subgroup. 
From the dynamics of the Dp-brane point view, if
we had put $N_f$ Dp and anti-Dp branes at two different  positions on  $S^1$  
whose other ends are stuck to the boundary of the background geometry then 
this corresponds to a situation with $U(N_f)\times U(N_f)$ global symmetry 
seen on the $N_c$ number of coincident D4-branes.  
From the study of the action of the flavor branes, which is governed by 
DBI and CS action,  one gets a
configuration in which the flavor branes are not anymore attached to $S^1$ i.e.not sitting at $u_{0}$ but are rather taking a $U$-shaped configuration which is at a ${\bar u}_0$ distance away from the core of the $N_c$ D4-branes. 
For this kind of configuration it means 
that the global symmetry group is now broken down to its diagonal subgroup 
$U(N_f)$.       
The topology of the flavor brane configurations  is that of a $U$-shaped 
configuration 
in which the Dp-branes and anti-Dp branes are joined together at some radial 
coordinate ${\bar u}_0$. This ${\bar u}_0$ can be greater or equal to $u_{0}$ 
as the later one 
corresponds to the minimum value of radial coordinate. Among these   
 U-shaped configurations,  the one that has highest energy when 
${\bar u}_0=u_0$  and the energy starts to decreases when ${\bar u}_0$ 
deviates away from  $u_0$,  at zero temperature. So, there will be breakdown of
 chiral symmetry  at zero temperature due to the choice of our brane embeddings.
Hence, we see
that the study of chiral symmetry in the confining phase yields the break 
down of the chiral symmetry group. i.e. the chiral symmetry breaking phase
stays in the the confining phase.    

If we Wick rotate  this background geometry then  we   generate  a 
black hole solution at a  temperature T and it breaks confinement. 
Now probing again this background by a bunch of coincident $N_f$ Dp-branes 
and anti-Dp branes, as previously,  we see that there is a restoration of 
chiral symmetry group. In this case there arises two distinct configurations: 
one whose topology is that of a U-shaped configuration and the other is 
two straight parallel branes sitting at two different point on a circle.  
So, we can say  depending on the value of T we have a nice phase diagram. 
In the current 
scenario we can have  two phase transitions, one is associated to 
confinement-deconfinement transition solely between  the zero temperature
background  and the black hole  background. The second is  the 
chiral symmetry breaking or 
restoration as seen through the flavor branes. These different phases can be 
combined among themselves as Confining (C) + Chiral symmetry braking 
($\chi_{sb}$), deconfinement (DC)+($\chi_{sb}$), DC+chiral symmetry 
restoration ($\chi_{sr}$) and finally C + $\chi_{sr}$. In the above 
D4/Dp/${\bar {Dp}}$ model there is no   C+ $\chi_{sr}$ phase in which the 
system can stay, which supports (\cite{cw}) in the sense of large $N_c$ limit. 

The confinement-deconfinement transition happens at a temperature 
$T=1/(2\pi R_{\tau})$, where $R_{\tau}$ 
is the radius of $S^1$ and the chiral symmetry breaking-restoration happens at 
T=c/L, where L is the asymptotic distance of separation between Dp and 
anti-Dp  branes and c is numerical factor, which in our case is less 
than one. 
From this it is interesting to note that for $L=2\pi c R_{\tau}$ one can't 
distinguish  the phases, whereas for L staying a bit above 
 $L=2\pi c R_{\tau}$   the DC and $\chi_{sr}$ happen together and for L 
staying from 0 to a bit less than $L=2\pi c R_{\tau}$ the dominated phase is 
that of  C+$\chi_{sb}$, DC+$\chi_{sb}$. The  C+$\chi_{sb}$, DC+$\chi_{sb}$ 
phases can be separated by the temperature T of the background. For T less than 
$1/(2\pi R_{\tau})$ the dominating phase is  C+$\chi_{sb}$. When the temperature
is above $1/(2\pi R_{\tau})$ but less then T=c/L then the preferred phase is 
 DC+$\chi_{sb}$. Hence there is a clear distinction between 
the C+$\chi_{sb}$, DC+$\chi_{sb}$ and DC+$\chi_{sr}$ phases.

Another result that follows from this study is that 
the asymptotic distance of separation, L,   between the quarks increases 
as one moves deep into the bulk at zero temperature case. At finite temperature
the distance between the quarks do not vary much but  as one approaches
close to the horizon the distance starts to decrease very fast. It is in this
region, which is very deep in the bulk, the distance changes very fast both 
at zero and finite temperature.

The scale of confinement depends on the parameters of geometry, 
for $N_c$ color D4-branes it depends   on $g_s N_c$ and
on the minimum value of the radial coordinate $u_0$ by eq.(\ref{tension_f1}). 
The chiral 
symmetry breaking scale  depends on the expectation value of fermion bilinear.
If we take it roughly as the distance of separation between the lowest point of
the U-shaped configuration to the minimum value of the radial coordinate. Then
one can distinguish these two scales. More interestingly one can make the 
chiral symmetry braking as high as possible by just moving  up this lowest 
point towards the boundary.  Explicitly, the  mass , in units of 
$\alpha^{'}=1$, is
\be
m=\f{1}{2\pi}\int^{{\bar u}_0}_{u_0} du \sqrt{-g_{00}g_{uu}}=u_0[0.137-\f{1}{6\pi} Beta[l^3, 2/3,1/2]]
\ee
where $0\leq l\leq 1$. 
From eq.(\ref{tension_f1}) and using the figures (\ref{fig_n_7_ltp}),(\ref{fig_n_8_ltp}) and (\ref{fig_n_9_ltp}), we can rewrite the 
tension of fundamental string as $T_{F1} < \f{u_0}{2\pi L_{ltp}}$ for 
some value of $l$, where $L_{ltp}$ is the asymptotic separation between the
flavor branes in the zero temperature phase.

The plan of the paper is: in section 2, we shall give the brane configuration
and the limit in which we are going to work. In section 3, the background 
geometry and in section 4, the chiral symmetry breaking in the zero temperature
phase and the associated NG boson and the puzzle. The puzzle is that 
for a specific kind of brane configuration, the CS action becomes important
to study the fluctuation associated to gauge fields and the massless NG boson.
But unfortunately we do not see NG boson for some specific cases, 
which is very puzzling. 
In section 5, we shall study the 
chiral symmetry restoration at finite temperature. 
In section 6,  we shall try to understand the chiral symmetry breaking/restoration in a different background and confinement-deconfinement transitions.
In particular by  introducing another electric charge to the near horizon limit
 of $N_c$ D4-branes by the charging up procedure used in black hole physics. 
The phase transition point in the phase diagram is 
studied with respect to the temperature T and the charge, which is related to 
the boost parameter.
So, the confinement-deconfinement transition is between the electrically 
charged non-confining black hole  background and the confining background 
and the parameter that describes this transition are: boost parameter and 
$\f{\beta}{2\pi R_{\tau}}$. In this case we see that when the boost parameter
takes a value  above 0.064 then the system stays in the confining phase 
irrespective of the value of  $\f{\beta}{2\pi R_{\tau}}$.
The chiral symmetry breaking/restoration  
transition is achieved with both the temperature and the boost parameter. For 
any finite boost we do see there is a chiral symmetry breaking and restoration.
It means we see the appearance of both the confining phase and the chiral
symmetry restored phase together, which is very confusing. We conclude in
section 8. In section 9, we have studied chiral symmetry breaking/restoration 
and 
massless NG boson for $QCD_4$ case in detail and in the last section we 
have worked out the  details of the inclusion of the electric field in the  
in the D4 background by uplifting the solution to 11 dimensional theory, then
boosting the solution and again reducing to generate the new solution.

\section{The brane configuration}

To study the chiral symmetry breaking,  we are considering  an intersecting  
brane configuration in Type IIA theory. The way we shall 
proceed is by considering a bunch of coincident $N_c$ D4-branes wrapped around
the 4th directions with radius $R_{\tau}$ and extended
along (0123) directions. 
After taking the  low energy limit we see the 
background  is 
magnetically charged under a 4-form antisymmetric field.  This background 
shows confining behavior. 
Let us add matter fields in the (bi)-fundamental representation
to this via probe-brane approximation. 

The kind of branes that we shall add are of the coincident $N_f$ Dp-brane and 
$N_f$ anti-Dp branes stuck at two different points along the 4th
direction and also are separated along some other spatial directions. 
The configuration looks as\\

\begin{tabular}{|l|l|l|l|l|l|l|l|l|l|l|l|c|}
\hline
Branes(p)        &0&1&2&3&4&5&6&7&8&9&$QCD_n$&\\
\hline
D4:              &x&x&x&x&x& & & & & &      &\\
\hline\hline
D4,${\bar{D4}}$: &x&x&x&x& &x& & & & &$QCD_4$&\\
D6,${\bar{D6}}$: &x&x&x&x& &x& & &x&x&$QCD_4$&\\
D8,${\bar{D8}}$: &x&x&x&x& &x&x&x&x&x&$QCD_4$&\\
\hline\hline
D4,${\bar{D4}}$: &x&x&x& & &x& & & &x&$QCD_3$&\\
D6,${\bar{D6}}$: &x&x&x& & &x& &x&x&x&$QCD_3$&\\
\hline\hline
D2,${\bar{D2}}$: &x&x& & & &x& & & & &$QCD_2$&\\
D4,${\bar{D4}}$: &x&x& & & &x& & &x&x&$QCD_2$&\\
D6,${\bar{D6}}$: &x&x& & & &x&x&x&x&x&$QCD_2$&\\
\hline\hline
\end{tabular}\\
\be
\label{brane_conf}
\ee
The 4th direction is  compact, 5th direction is the radial direction 
and 6, 7, 8 and 9th direction makes an $S^4$.    
The $S0(1,9)$ symmetry is broken by the color $N_c$ D4-branes to 
S0(1,4)$\times$ S0(5). The 
flavor $N_f$ Dp-branes and anti-Dp branes also break S0(1,9) down to its 
subgroup. Let $n$ be the number of directions common to both color and flavor
branes. Then the field theory, before taking the low energy limit, is 
n-dimensional. Which we have written as $QCD_n$. The symmetry that is 
preserved both by the color and flavor brane in general has a structure
that of  $S0(1,n-1)\times S0(p-n+1)$, for n=2,3,4. There could possibly be
another group and its not difficult to figure out its structure. For an example
the $D2-{\bar{D2}}$ for $QCD_2$ case the S0(1,9) symmetry is broken to 
$S0(1,1)_{01}\times S0(2)_{23}\times S0(4)_{6789}$. The flavor branes are 
separated by a distance $L$ and the directions along which they are separated
can be figured out from (\ref{brane_conf}), for example the $D2-{\bar{D2}}$ 
for 
$QCD_2$ case the flavor branes are separated along (2346789) directions. 
There is one more important point to note is that there is no chirality for 
$QCD_3$ case, in the sense of having Weyl fermions. 

This brane intersection   in the low energy limit gives various fields
that arises from p-p, p-${\bar{p}}$, ${\bar{p}}-{\bar{p}}$, 4-p, 4-${\bar{p}}$
strings. As suggested in \cite{ahjk} in the limit of $N_c\rightarrow\infty, 
g_s\rightarrow 0$ and fixing $g_sN_c, N_f$ with $N_f < < N_c, L>>l_s$, the 
coupling of 
p-p,  p-${\bar{p}}$, and ${\bar{p}}-{\bar{p}}$, strings goes to zero and 
becomes non-dynamical sources. Hence, the degrees of freedom in the low 
energy limit and in   the above limiting case is described by the 4-4, 4-p and
4-${\bar{p}}$ strings. The gauge symmetry of the flavor brane $U(N_f)\times
U(N_f)$ becomes a global symmetry of the n dimensional theory at the 
intersection of the brane. The fermions that appear from the 4-p intersection
transform as $({\bar N}_f,1)$ of the global symmetry and that of 4-${\bar{p}}$ 
intersection as $(1,{\bar N}_f)$. Both fermions transform in the fundamental 
$N_c$ representation of the color group.

The 't Hooft coupling is dimensionful and has got unit length dimension. The
theory becomes weakly coupled in the $L >> \lambda$ and strongly coupled 
in the $\lambda >> L$ limit. Before taking the 4th direction as compact, 
in the near horizon limit the string coupling goes to infinity in UV. 
It means this kind of 
theories are non-normalisable. 
The UV completed theory \cite{imsy} is described by a six
dimensional  (0,2) CFT   compactified on a circle of
radius $g_s\sqrt{\alpha^{'}}$. In the weak coupling limit its been shown that
the single gluon exchange dominates over multiple gluons between the 
fermions coming from the intersection of color and flavor branes, for details
see\cite{ahjk}. We shall study this intersection in the opposite limit i.e.
in the strongly coupled limit.

\section{The background}

In the strong coupling $\lambda>> L$ limit, it means 
the easiest way to study the system in this regime is by bringing the 
flavor branes closer. The background geometry that we are considering is 
that of a near horizon limit
of a bunch of $N_c$ coincident D4 branes which acts as the source for the 
4-form 
magnetic field. The topology of the geometry is $R^{1,3}\times S^1\times S^4\times R^1$. The geometry along $R^{1,3}$ is  flat but the presence 
of  magnetic field, which has got fluxes along $S^4$, back reacts to  
create warping on the 
10-dimensional geometry. Hence it makes conformally flat along $R^{1,3}$. 
Moreover, the size of $S^1$ vanishes at the
lower range of radial coordinate and there is non-trivial dilaton. Writing 
down the geometry explicitly in string frame
\beqn
\label{con_d4}
& &ds^2=(u/R)^{3/2}(\eta_{\mu\nu} dx^{\mu}dx^{\nu}+f(u)d\tau^2)+(R/u)^{3/2}(\frac{du^2}{f(u)}+u^2d\Omega^2_4),\nn \\
& &e^{\phi}=g_s (u/R)^{3/4},\quad F_4=\frac{2\pi N_c}{V_4}\epsilon_4, \quad f(u)=1-(\frac{u_0}{u})^3.
\eeqn
The D4-branes are extended along $x^{\mu}, \tau$ directions and smeared over $S^4$. $\epsilon_4, V_4$ are the unit volume form and volume of $S^4$ 
respectively. From the expression to $f(u)$ it follows that the radial 
coordinate stays from $u_0$ to $\infty$. Since the $\tau$ direction is compact
one can compactify the fermions to satisfy anti-periodic boundary
 condition \cite{ew}, so as to break supersymmetry, before taking the 
near horizon 
limit. In what follows we shall also break 
supersymmetry explicitly by taking a situation where a brane and anti-brane has
been placed on the $S^1$. Another
 important point to note that the periodicity of the $S^1$ is not $2\pi$ but
rather 
\be
\tau ~~~\sim ~~~ \tau+\delta\tau, \quad \delta\tau=\f{4\pi}{3}
(\f{R^3}{u_0})^{1/2}.
\ee

Given this background one can convince one self that this is a confining 
background with the tension of the flux tube evaluated at the minimum value to
the radial coordinate is 
\be
\label{tension_f1}
T_{F1}=\f{1}{2\pi\alpha^{'}}(\f{u_0}{R})^{3/2},\quad R^3=\pi g_s N_c 
(\alpha^{'})^{3/2}.
\ee 

With this background and following the prescription of \cite{ss}, which is one
of the way to understand chiral symmetry breaking and restoration, is to 
probe this background with a bunch of $N_f$ coincident Dp-branes localized 
on some 
point on $S^1$ and putting another bunch of $N_f$ anti-Dp-branes on some 
other point of $S^1$. Now the global symmetry as seen from the D4-brane 
point of view is $U(N_f)\times U(N_f)$, which means the two set of branes do 
not touch each other and breaking of this global symmetry down to its diagonal
subgroup $U(N_f)$ corresponds to joining the branes with the antibranes.
It means the break down of chiral symmetry. Here there
arises a question what if we had taken $N_f$ Dp-branes and ${\tilde{N_f}}$ 
anti-Dp branes with $N_f$ not necessarily same as ${\tilde{N_f}}$? The 
possibilities are  there could be a configuration with  unbroken product 
global symmetry, i.e. $U(N_f) \times U({\tilde{N_f}})\rightarrow U(n_f)\times 
U({\tilde{n}_f})$ with $N_f, {\tilde N}_f > n_f, {\tilde n}_f$, which means 
the chiral symmetry is unbroken. But, if  $U(N_f )\times U({\tilde{N_f}})
\rightarrow U(n_f)$ then there is a signature of chiral symmetry breaking.
For $N_f$ is not same as  ${\tilde{N_f}}$, it seems we can't study that
in this near horizon limit. 
For our purpose  we shall be taking the case where we have got equal 
number of branes and anti-branes. This is the setting which has been studied 
recently in \cite{ahjk} in the weak coupling limit  and \cite{asy}, \cite{ps}
in the finite temperature case for p=8 case.

\section{Strongly coupled action: in the low temperature phase}

\begin{figure}[htb]
\includegraphics{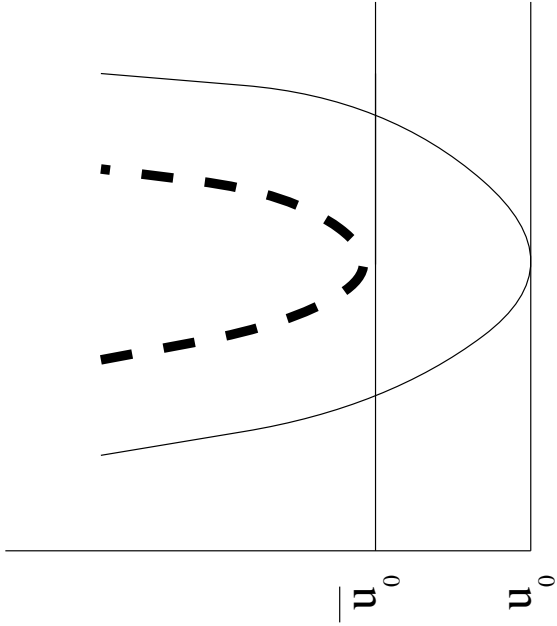}
\caption{The  two possible U-shaped configurations are  dotted-thick line 
corresponds to the non-zero momentum whereas the thin-solid line corresponds
to zero momentum $||$-configurations, in the 
zero temperature phase. }
\label{fig_n_0_ltp}
\end{figure}

The dynamics of flavor branes in the near horizon limit of $N_c$ D4 branes are
described by the DBI and CS action. For the flavor branes that we are 
considering, its not difficult to convince that there won't be any CS term in 
the action. To compute the DBI part, we need to compute the induced metric on
the worldvolume of the flavor Dp-brane. The induced metric is
\be
ds^2_{Dp}=(\f{u}{R})^{(3/2)}[-dt^2+dx^2_{n-1}]+(\f{u}{R})^{(3/2)}[f\tau^{'2}+(R/u)^3 \f{1}{f}]du^2+R^{(3/2)} u^{(1/2)}d\Omega^2_{p-n},
\ee
where $0 \leq p-n \leq 4$ and we have fixed the p-branes at $\theta_i=\pi/2$
for simplicity, wherever it required, in particular we use the metric of $S^4$
as $d\Omega^2_4=d\theta^2_1+sin^2\theta_1 d\theta^2_2+
sin^2\theta_1 sin^2\theta_2 (d\theta^2_3+sin^2\theta_3 d\phi^2)$. 
The DBI part of the action now becomes
\be
S_{Dp}(QCD_n;LTP)=-\f{T_p}{g_s} V_{p-n} V_n R^{\f{3}{4}(p-2n)}\int du u^{(p+2n)/4}\sqrt{f(\f{d\tau}{du})^2+(R/u)^3\f{1}{f}}.
\ee
Since the action do not depend on $\tau(u)$ explicitly, 
implies the conjugate momentum 
associated to it is constant. Assuming the boundary condition 
$u(\tau=0)\rightarrow {\bar{u}}_0$ with $u^{'}(\tau=0)=0$ gives
\be
u^{'}=f (\f{u}{R})^{(3/2)} [\f{ u^{(p+2n)/2} f}{{\bar u}^{(p+2n)/2}_0 f_0}-1]^{(1/2)}.
\ee  
Integrating this one can find the most general solution. The choice of our 
boundary condition tells us that this configuration is a symmetric U-shaped 
curve in (U,$\tau$) plane. There is a special U-shaped configuration for 
 which it  goes all the way to the bottom of the 
$\tau$ circle, but by a slight abuse of notation we shall denote it as 
  $||$, for which 
the momentum vanishes i.e. $\tau^{'}=0$. This means 
$u^{'}=\f{1}{\tau^{'}}=\infty$ and it happens
when ${\bar u}_0\rightarrow u_0$. The function $f_0=1-(\f{u_0}{{\bar u}_0})^3$.
In both the configuration the flavor branes are located at 
$\tau=\pm \f{L}{2}$, asymptotically. 
In this case the chiral symmetry is broken explicitly before making any 
calculation 
because we have put flavor branes at two different points on the $\tau$ circle 
and this circle shrinks to zero size. However, to figure out which of the 
U-shaped configuration has got the lower energy, i.e. is it the one which
stays close to the shrinking $\tau$ circle or the one which stays away from
it. These two different configurations are plotted in figure 
(\ref{fig_n_0_ltp}).

Computing the value of the action for these two types of configurations and 
taking their difference one end up with
\beqn
S(U)-S(||)&=&-[~]\Bigg( \int^1_0 dz z^{-\f{p+2n+10}{12}} \bigg[ \f{1}{\sqrt{1-l^3 z-(1-l^3) z^{\f{p+2n}{6}}}}-\f{1}{\sqrt{1-l^3 z}}\bigg]-\nn \\& &\int^{\f{1}{l^3}}_1 dz \f{z^{-\f{p+2n+10}{12}}}{\sqrt{1-l^3 z}}\Bigg), 
\eeqn 
where $l=\f{u_0}{{\bar u}_0}$. In general it is very difficult to find the 
difference between the two configurations analytically. So, we shall do it 
numerically by using Mathematica package. The parameter $l$ can take maximum
value 1, when the turning point ${\bar u}_0$ coincides with $u_0$, the minimum
value to $u$, and it takes minimum value when the turning point approaches the
boundary i.e. $0\leq l \leq 1$. We have not given the expression to $[~]$. This
factor is  positive and comes from the prefactor that multiplies the DBI 
action, integration over the compact and non-compact directions and doing some
change of variables. we find it is given by $[~]=\f{T_p}{3g_s} V_{p-n}V_n R^{\f{3(p-2n+2)}{4}}{\bar u}^{\f{p+2n-2}{4}}_0$.  

\begin{figure}[htb]
\includegraphics{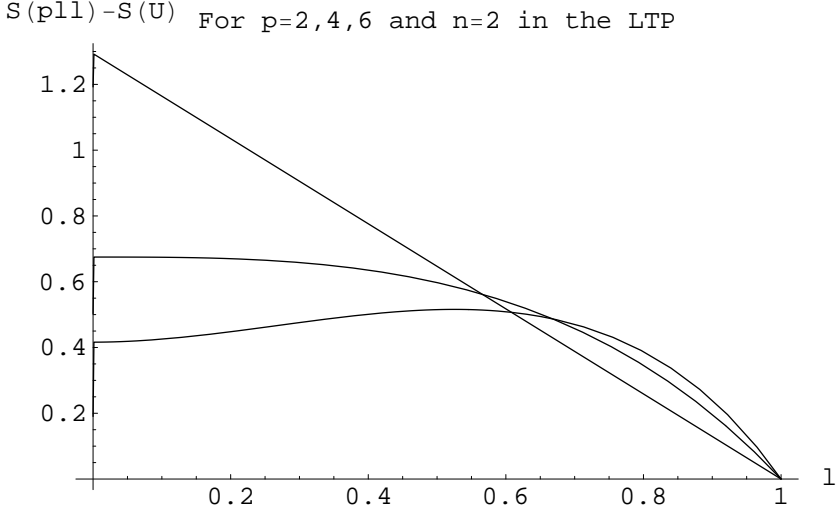}
\caption{The difference between the actions, $S_{Dp}(||)-S_{Dp}(U)$,  is plotted against $l=\f{u_0}{{\bar u}_0}$ for $QCD_2$ case and for 
D2-${\bar D2}$,D4-${\bar D4}$ and $D6-{\bar D6}$ branes in the zero temperature
case.}
\label{fig_n_2_ltp}
\end{figure}

\begin{figure}[htb]
\includegraphics{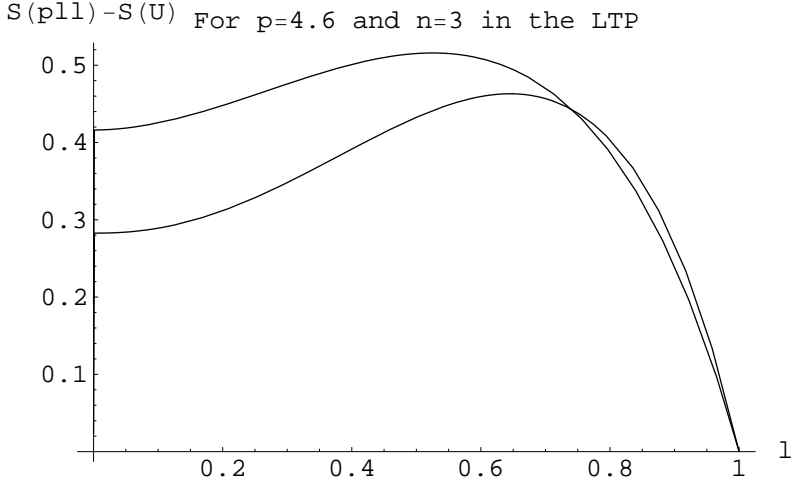}
\caption{The difference between the straight and U-shaped action, $S_{Dp}(||)-S_{Dp}(U)$,  is plotted against $l=\f{u_0}{{\bar u}_0}$ for $QCD_3$ case for 
 D4-${\bar D4}$ and D6-${\bar D6}$ branes in the zero temperature case.}
\label{fig_n_3_ltp}
\end{figure}

The difference between these two U-shaped configurations, 
$S(||)-S(U)$ vs $l$, is plotted in figure (\ref{fig_n_2_ltp}) 
and (\ref{fig_n_3_ltp}). 

From these plots we see that  it is the action of the U-shaped  
configuration which stays 
away from the bottom of the $\tau$ circle ,$||$,  
 dominates over the one which stays close to the bottom of the $\tau$
circle configuration. 
Stating it in terms of the energy, we find that $||$-configuration 
has got more energy, as in     the Minkowski
signature the action is the negative of energy. Hence we find that at 
 zero temperature the global symmetry $U(N_f)\times U(N_f)$ is
broken to its diagonal subgroup $U(N_f)$, which is described by this U-shaped
configuration. In this case the breakdown of the chiral symmetry is guaranteed
due to the presence of the shrinking $\tau$-circle. However, in the cases when 
there is no shrinking of $\tau$-circle, still one can achieve  chiral symmetry 
breaking as in \cite{ahjk}. Which means at zero temperature there is  
break down of chiral symmetry. 

\subsection{NG Boson}

It is known that the break down of a global symmetry would mean the presence of
a massless Nambu-Goldstone boson in the spectrum. The way to see this massless scalar in the 
break down of the chiral symmetry is to look for the fluctuation to the 
U(1) gauge
potential that live on the worldvolume of flavor brane by \cite{ss}. Let us
go through the procedure and see whether there appears any NG boson in the 
spectrum. 

The  fluctuation in the gauge potential to the DBI action of a Dp brane can be 
written as
\be
S_{DBI}={T_p}\int e^{-\phi}\sqrt{-det[g]} \bigg( \f{1}{4}[g]^{ab}F_{bc}[g]^{cd}F_{da}\bigg)+{\cal O}(F^3),
\ee 
where we have used the notation $[g]$ as the pullback of the metric onto the 
worldvolume of the p-brane. Exciting the fields only along the S0(1,n-1) 
and the radial direction sets the DBI action as
\be
S_{DBI}={T_p}\int e^{-\phi}\sqrt{-det[g]} \f{1}{4}\bigg([g]^{\mu\nu}F_{\nu\rho}[g]^{\rho\sigma}F_{\sigma\mu}+2[g]^{\mu\nu}F_{\nu u}[g]^{uu}F_{u\mu}\bigg),
\ee
where $\mu,\nu=0,1,\ldots,n-1$. There could in general be the fluctuation to 
gauge potential in CS action and for our purpose it is easy to see that 
only for
D6-${\bar D6}$ in the $QCD_2$ and $ QCD_3$ case there can be a term in 
the CS action which supports fluctuations to quadratic order. 
For a D6-brane the action is
\be
S_{CS}=\f{\mu_6}{2}\int [C_3]\wedge F^2+{\cal O}(F^3)=\f{3}{4}\mu_6 k N_c\int F^2.
\ee   
For the $QCD_3$ case this  indeed is the form of   CS action, whereas
for the $QCD_2$ case 
\be
S_{CS}=\f{\mu_6}{2}\int [F_4]\wedge A\wedge F+{\cal O}(F^3)=\mu_6N_c\pi\int A\wedge F
\ee
this form of the action is useful. k is a constant that appear in the three 
form antisymmetric tensor. Note the branes have been fixed at 
$\theta_i=\pi/2$. Let us rewrite the flavor brane action for $QCD_2$ case as
\beqn
S_{D6}(QCD_2)&=&T_6\int  e^{-\phi}\sqrt{-det[g]} \f{1}{4}\bigg([g]^{\mu\nu}F_{\nu\rho}[g]^{\rho\sigma}F_{\sigma\mu}+2[g]^{\mu\nu}F_{\nu u}[g]^{uu}F_{u\mu}\bigg)+
\nn \\& &
\f{b}{2}\int
d^2x du \epsilon^{\rho\sigma}(A_uF_{\rho\sigma}+2 A_{\sigma}F_{\sigma u}),
\eeqn
where $b=\mu_6 N_c\pi$ and b  vanishes  for other branes. Similarly, for 
$QCD_3$ case the action for flavor D6-brane is
\beqn
S_{D6}(QCD_3)&=&T_6\int  e^{-\phi}\sqrt{-det[g]} \f{1}{4}\bigg([g]^{\mu\nu}F_{\nu\rho}[g]^{\rho\sigma}F_{\sigma\mu}+2[g]^{\mu\nu}F_{\nu u}[g]^{uu}F_{u\mu}\bigg)+\nn \\& &
a \int
d^3x du \epsilon^{\mu\nu\rho}(F_{\mu\nu}F_{\rho u}),
\eeqn
with $a=\f{3}{4}\mu_6 k N_c$ and a vanishes for other branes. So, the 
fluctuated action 
of the flavor brane  is described by the sum of DBI and CS action. 
Computing the DBI part of the action for our purpose we find it can be 
rewritten as 
\be
S_{Dp}(QCD_n;LTP)=-[~]\int d^nx du [\f{K_1(u)}{4} F_{\mu\nu} F^{\mu\nu}+\f{K_2(u)}{2}\eta^{\mu\nu}F_{\mu u} F_{\nu u}],
\ee
where 
\be
\f{K_1(u)}{4}=\f{R^{\f{3}{2}} u^{\f{p+2n-9}{2}}}{\sqrt{u^{\f{p+2n}{2}} f-{\bar u}^{\f{p+2n}{2}}_0 f_0}},\quad \f{K_2(u)}{4}=R^{-\f{3}{2}} u^{-\f{3}{2}}\sqrt{u^{\f{p+2n}{2}} f-{\bar u}^{\f{p+2n}{2}}_0 f_0}.
\ee
As usual $[~]=\f{T_p}{4g_s} V_{p-n} R^{\f{p-2n+4}{4}}$ contains some 
 over all positive constants. Expanding the gauge 
potential as
\beqn
& &A_{\mu}(x^{\mu},u)=\sum_{n=1} B^{(n)}_{\mu} (x^{\mu})\psi_n(u),\quad A_{u}(x^{\mu},u)= \sum_{n=0} \pi^{(n)} (x^{\mu}) \phi_n(u)\nn \\& & 
F_{\mu\nu}=\sum_nF^{(n)}_{\mu\nu}(x^{\mu})\psi_n(u),\quad F_{\mu u}=\sum_n (\partial_{\mu}\pi^{(n)}(x)\phi_n(u)-B^{(n)}_{\mu}(x)\partial_u\psi_n(u)),
\eeqn
where $F^{(n)}_{\mu\nu}=
\partial_{\mu}B^{(n)}_{\nu}-\partial_{\nu}B^{(n)}_{\mu}$. Let us impose proper
eigenvalue equation for $\psi_n$ and normalization conditions for the mode 
$\psi_n$ and $\phi_n$. The conditions are
\beqn
\label{nor_eigen}
& &-\partial_u(K_2 \partial_u\psi_n)=K_1\lambda_n\psi_n,\quad [~]\int du K_1 \psi_n\psi_m=\delta_{nm}\nn \\& &
[~]\int du K_2\phi_m\phi_n=\delta_{mn}, \quad [~]\int du K_2 \partial_u \psi_n\partial_u\psi_m=\lambda_n\delta_{nm}=m^2_n\delta_{nm}.
\eeqn
From equation \ref{nor_eigen} there arises two relations
\be
\phi_n=\sqrt{\f{K_1}{K_2}}\psi_n, \quad \phi_n=\f{1}{m_n}\partial_u\psi_n.
\ee

In order to evaluate the fluctuated action we need one more information about 
the zero mode of $\phi_0$. From the equation of motion to the DBI and CS 
action  we find that 
\be
\phi_0=c_1/K_2(u),
\ee 
where $c_1$ is a constant. The asymptotic behavior of $K_1(u)$ and 
$K_2(u)$ are
\be
K_1(u)\rightarrow  4 R^{(3/2)}u^{\f{p+2n-18}{4}},\quad K_2(u)\rightarrow 4R^{-(3/2)}u^{\f{p+2n-6}{4}}.
\ee
We have defined the normalization condition for $\phi_n$ in such a way so as 
to generate the canonical kinetic term for $\pi^{(n)}$, which is true for the
zero mode of $A_u$ along the radial direction i.e. $\phi_0$ obeys
\be
\label{nor_phi_0}
[~]\int du K_2 \phi^2_0=1
\ee
Now the LHS of equation (\ref{nor_phi_0}) in the asymptotic limit of $K_2$
becomes
\be
\sim [] \int du \f{1}{u^{\f{p+2n-6}{4}}}. 
\ee

The range of p+2n is $0\leq \f{p+2n-6}{4} \leq \f{5}{2}$, which means that 
the above
integral is not normalisable in contrast to our  assumption earlier. In fact 
for some flavor Dp-branes it diverges linearly and for some other 
logarithmically. This 
happens when P+2n=6 and p+2n=10 respectively.So, for $QCD_2$ case and for 
flavor D2 brane it diverges linearly and for flavor D6-branes it diverges logarithmically, whereas for $QCD_3$ case, for flavor D4-branes it diverges 
logarithmically. For other cases the mode $\phi_0$ is normalisable. 

Does this mean the absence of a normalisable zero mode  $\phi_0$ for these 
flavor brane cases do not have any chiral symmetry breaking? But we saw 
the chiral symmetry breaking in the low temperature phase explicitly. So, 
then where is the NG boson? 

The fluctuated actions for $QCD_2$ and $QCD_3$ are
\beqn
S(QCD_3)&=&-\int d^3x[\sum_{n=1}( \f{1}{4}F^{(n)}_{\mu\nu}F^{(n)\mu\nu}+
\f{1}{2}m^2_n B^{(n)}_{\mu}B^{(n)\mu})+\f{1}{2}\partial_{\mu}\pi^{(0)}\partial^{\mu}\pi^{(0)}]-\nn \\& &
a \sum_{n,m=1}M^2_{nm}\int d^3x\epsilon^{\mu\nu\rho}F^{(n)}_{\mu\nu}B^{(m)}_{\rho}+a \sum_{n=1}M^2_n\int d^3 x\epsilon^{\mu\nu\rho}F^{(n)}_{\mu\nu}\partial_{\rho}\pi^{(0)}\nn \\
S(QCD_2)&=&-\int d^3x[\sum_{n=1}( \f{1}{4}F^{(n)}_{\mu\nu}F^{(n)\mu\nu}+
\f{1}{2}m^2_n B^{(n)}_{\mu}B^{(n)\mu})+\f{1}{2}\partial_{\mu}\pi^{(0)}\partial^{\mu}\pi^{(0)}]-\nn \\& &
\f{b}{2}\sum_{n,m=1}\int d^2 x\epsilon^{\rho\sigma}[M^2_n(\pi^{(0)}F^{(n)}_{\rho\sigma}+2(B^{(n)}_{\rho}+\f{1}{m_n}\partial_{\rho}\pi^{(n)})\partial_{\sigma}\pi^{(0)})\nn \\& &
M^2_{mn}(\f{1}{m_n}\pi^{(n)}F^{(m)}_{\rho\sigma}-2B^{(n)}_{\sigma}(B^{(m)}_{\rho}+\f{1}{m_n}\partial_{\rho}\pi^{(m)}))],
\eeqn
where
\be
\int du \psi_n\partial_u\psi_m=M^2_{nm}=-M^2_{mn},\quad \int du \phi_0 \psi_n=M^2_n.
\ee

From this we do get massless boson for $QCD_3$  with D6-brane and anti-D6 
brane case but the kinetic term for 
$\pi^{(0)}$ is not canonically normalized. However we do not get any massless 
NG boson for $QCD_2$ case. For the $QCD_2$ with $D2-{\bar D2}$ flavor branes 
there is no CS term but the mode $\phi_0$ is not normalized. Hence the 
$\pi^{(0)}$
that appear in the DBI do not make sense and the  same is true for 
$D4-{\bar D4}$ brane
in the $QCD_3$ case. The $\pi^{(0)}$ field in these cases can't be 
interpreted as  the NG bosons.

Let us work in a different 
gauge choice, $A_u=0$. In this gauge choice,
$A_u\rightarrow A_u-\partial_u\Lambda$, 
we can take $\Lambda=\sum\f{1}{m_n}\pi^{(n)}\psi_n$. Then $A_{\mu}=\sum_n(B^{(n)}_{\mu}-\f{1}{m_n}\partial_{\mu}\pi^{(n)})\psi_n$. Computing the fluctuated 
action for $QCD_2$, for example, we find
\beqn
S(QCD_2)&=&-\int d^2x[ \f{1}{4}F^{(0)}_{\mu\nu}F^{(0)\mu\nu}+\sum_1(\f{1}{4}F^{(n)}_{\mu\nu}F^{(n)\mu\nu}+\f{1}{2}m^2_n B^{(n)}_{\mu} B^{(n)\mu})]-\nn \\& &
b\sum_{n,m} \int d^2x \epsilon^{\rho\sigma}M^2_{nm} B^{(n)}_{\rho} B^{(m)}_{\sigma},
\eeqn 
where we have absorbed that extra factor into $ B^{(n)}_{\mu}$ and the sum 
in the expansion of $A_{\mu}(x,u)$ in terms of modes  runs from 0 to $\infty$.  
Similarly one can compute for the $QCD_3$ case in this gauge. The result is
\beqn
S(QCD_3)&=&-\int d^3x [\f{1}{4}F^{(0)}_{\mu\nu}F^{(0)\mu\nu}+\sum_1(\f{1}{4}F^{(n)}_{\mu\nu}F^{(n)\mu\nu}+\f{1}{2}m^2_n B^{(n)}_{\mu} B^{(n)\mu})\-\nn \\& &
a \sum_{n,m} \int d^3x \epsilon^{\mu\nu\rho}M^2_{nm} F^{(n)}_{\mu\nu} B^{(m)}_{\rho} 
\eeqn

The conclusion 
is again as before, we do not see any massless NG boson for $QCD_2$ case. 

\section{Chiral symmetry at finite temperature}

The probe brane computation of a bunch of coincident Dp-brane in the 
supergravity limit with the background that of near horizon limit of 
a bunch coincident $N_c$ D4-branes at finite temperature is

 \beqn
\label{noncon_d4}
& &ds^2=(u/R)^{3/2}(f (u) dt^2+\sum^3_1\delta_{ij} dx^{i}dx^{j}+d\tau^2)+(R/u)^{3/2}(\frac{du^2}{f(u)}+u^2d\Omega^2_4),\nn \\
& &e^{\phi}=g_s (u/R)^{3/4},\quad F_4=\frac{2\pi N_c}{V_4}\epsilon_4, \quad f(u)=1-(\frac{u_T}{u})^3.
\eeqn

The periodicity of the Euclidean time direction is 
\be
\beta=\f{4\pi}{3} (\f{R^3}{u_T})^{(1/2)}.
\ee

The Euclidean action of probe Dp brane in this background is
\be
S_{Dp}(QCD_n)=\f{T_p}{g_s}V_{p-n}V_n R^{\f{3(p-2n)}{4}}\int du \sqrt{f}u^{\f{p+2n}{4}}\sqrt{\tau^{'2}+(1/f)(R/u)^3},
\ee
where the induced metric is
\be
ds^2_{Dp}=(\f{u}{R})^{(3/2)}[fdt^2+dx^2_{n-1}]+(\f{u}{R})^{(3/2)}[\tau^{'2}+(R/u)^3 \f{1}{f}]du^2+R^{(3/2)} u^{(1/2)}d\Omega^2_{p-n},
\ee
 and we have fixed the flavor brane at $\theta_i=\pi/2$, whenever required, 
for simplicity.
Since the action do not depend upon $\tau(u)$ explicitly, it means the corresponding momentum is conserved. As before, there are two independent 
configurations which solve
the equation of motion. One is when $\tau^{'}=0$ or its inverse $\f{du}{d\tau}$
diverges. This corresponds to a straight configuration and in this case we do
not break the global symmetry $U(N_f)\times U(N_f)$. To remind about this symmetry, we have taken Dp-brane and anti-Dp brane and put them at two different 
points on the compact $\tau=x^4$ circle. The second configuration is that of
a U-shaped solution. This solution is described by a curve in the $u,\tau$ 
plane and is
\be
u^{'}=\sqrt{f} (u/R)^{3/2}[\f{u^{\f{p+2n}{2}}f}{{\bar u}^{\f{p+2n}{2}}_0f_0}-1]^{1/2},
\ee 
where $f_0=1-(\f{u_T}{{\bar u}_0})^3$. The U-shaped brane configuration joins
 Dp-brane with anti-Dp brane  at ${\bar u}_0$.

In order to determine  the lowest energy configuration, we have to
find the difference between the actions and determine which configuration
is preferred from it. 
Substituting this velocity in the action for the U-shaped configuration, we 
find
\be
S(U)=[~] \int^{\infty}_{u_T} du \f{u^{\f{p+2n-6}{4}}\sqrt{f}}
{\sqrt{f-(\f{{\bar u}_0}{u})^{\f{p+2n}{2}}f_0}}.
\ee

Similarly the action for the straight Dp-brane and anti-Dp-brane is
\be
S(||)=[~] \int^{\infty}_{u_0} du  u^{\f{p+2n-6}{4}}.
\ee
 
\begin{figure}[htb]
\includegraphics{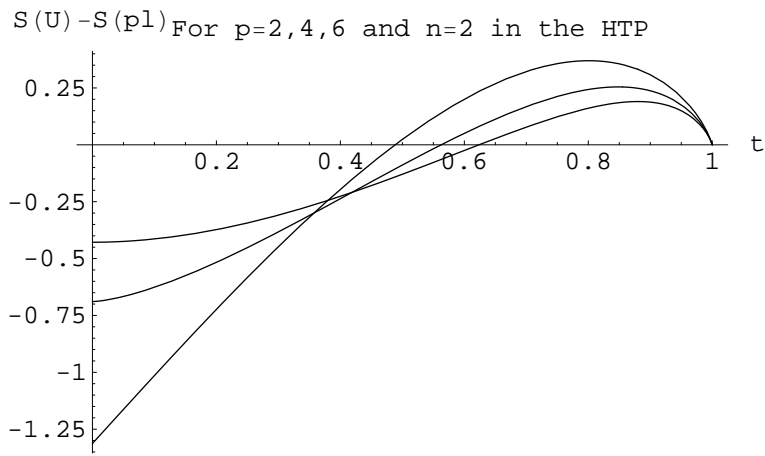}
\caption{The difference between the two possible solutions, $S_{Dp}(U)-S_{Dp}(||)$, is plotted against $t=\f{u_T}{{\bar u}_0}$ for $QCD_2$ case and for different flavor branes: D2-${\bar D2}$ ,D4-${\bar D4}$ and D6-${\bar D6}$, in the 
finite temperature phase. }
\label{fig_n_5_htp}
\end{figure}

\begin{figure}[htb]
\includegraphics{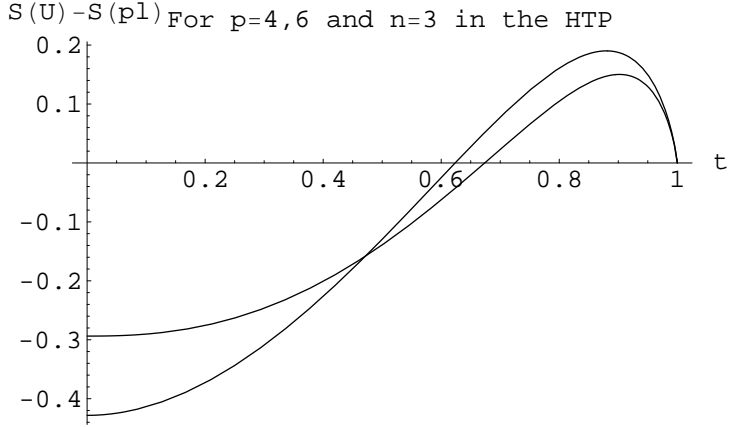}
\caption{The difference between the actions for the two possible solutions, 
$S_{Dp}(U)-S_{Dp}(||)$, is plotted against $t=\f{u_T}{{\bar u}_0}$ for $QCD_3$ and for flavor brane: D4-${\bar D4}$ and D6-${\bar D6}$ in the finite 
temperature case.}
\label{fig_n_6_htp}
\end{figure}

Note the integration for the straight configuration case can be written as 
a sum of two pieces one from 
$u_0$ to ${\bar u}_0$ and the other  from ${\bar u}_0$ to $\infty$. 
Doing some change of variables and finding the difference between the 
actions, we find
\beqn
S(U)-S(||)&=&[~]\Bigg(\int^1_0 dz z^{-\f{p+2n+10}{12}}\bigg(\sqrt{\f{1-t^3 z}{1-t^3 z-(1-t^3)z^{\f{p+2n}{6}}}} -1\bigg)-\nn \\& &\f{12}{p+2n-2} \bigg(1-t^{\f{p+2n-2}{4}} \bigg)  \Bigg),
\eeqn
where $0\leq t=\f{u_T}{{\bar u}_0}\leq 1$. Now, finding the difference is not 
easy, so instead we plot the difference using Mathematica package. The 
difference between the action is plotted separately for different value of $n$
in figures \ref{fig_n_5_htp} and \ref{fig_n_6_htp}.


Let the asymptotic distance between the flavor Dp-branes and anti-Dp branes 
be $L_{HTP}$ in the finite  temperature case and $L_{LTP}$ for the zero 
 temperature case. The expressions for them are

\be
L_{HTP}={\f{2}{3}}\sqrt{\f{R^3}{{\bar u}_0}}{\sqrt{1-t^3}} \int^1_0 dz 
\f{z^{\f{p+2n-10}{12}}}{\sqrt{1-t^3 z}\sqrt{1-t^3 z-(1-t^3)z^{\f{p+2n}{6}}}}
\ee
\be
L_{LTP}={\f{2}{3}}{\sqrt{\f{R^3}{{\bar u}_0}}}{\sqrt{1-l^3}} \int^1_0 dz 
\f{z^{\f{p+2n-10}{12}}}{(1-l^3 z)\sqrt{1-l^3 z-(1-l^3)z^{\f{p+2n}{6}}}}.
\ee
\begin{figure}[tb]
\includegraphics{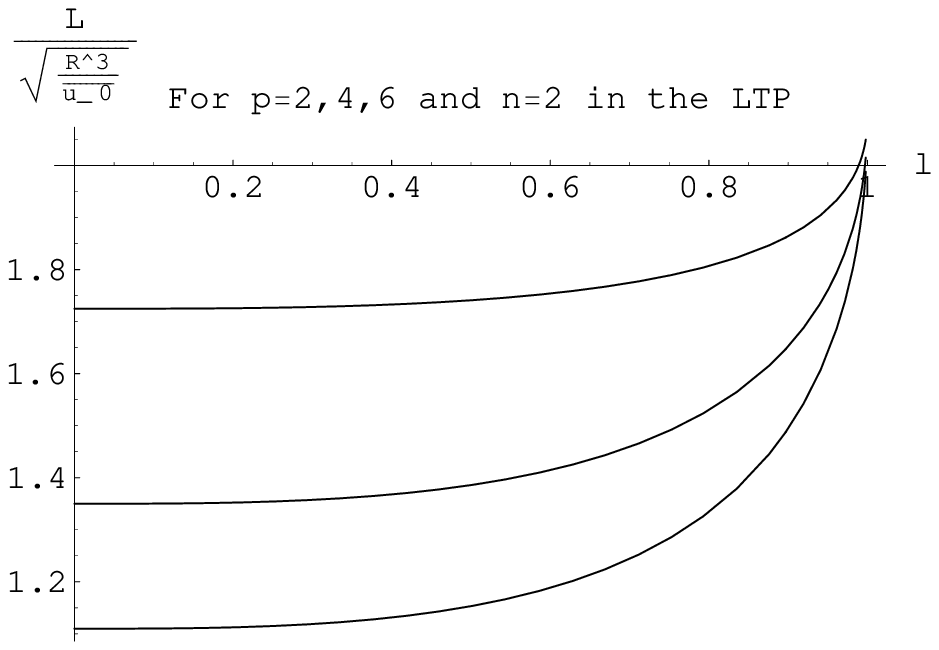}
\caption{$\f{L}{\sqrt{R^3/{\bar u}_0}}$ is plotted against $l=\f{u_0}{{\bar u}_0}$ for D2-${\bar D2}$, D4-${\bar D4}$ and D6-${\bar D6}$ flavor branes for the $QCD_2$ case in zero temperature phase.}
\label{fig_n_7_ltp}
\end{figure}
\begin{figure}[tb]
\includegraphics{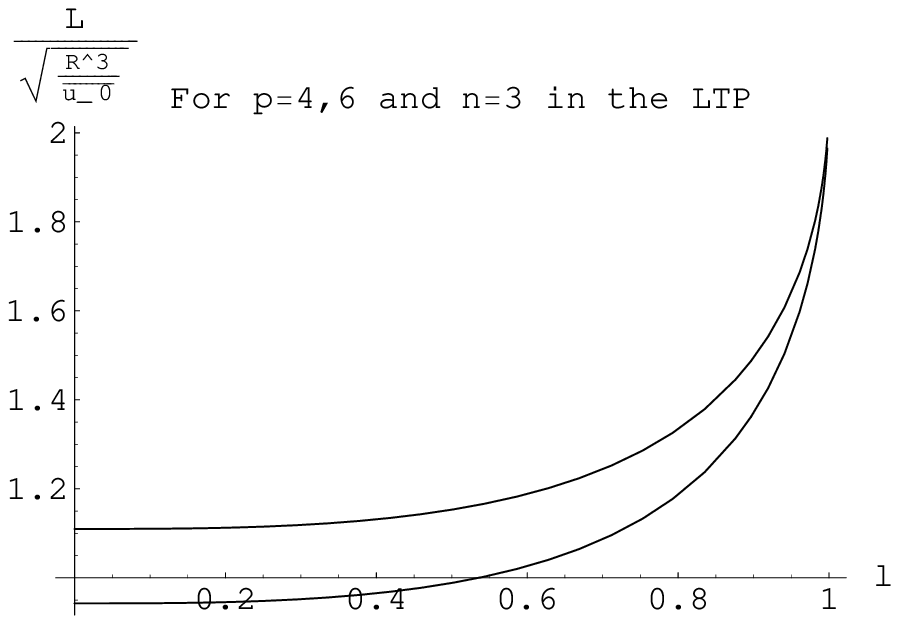}
\caption{$\f{L}{\sqrt{R^3/{\bar u}_0}}$ is plotted against $l=\f{u_0}{{\bar u}_0}$ for flavor branes: D4-${\bar D4}$ and D6-${\bar D6}$ for the $QCD_3$ case in zero temperature case.}
\label{fig_n_8_ltp}
\end{figure}
\begin{figure}[ht]
\includegraphics{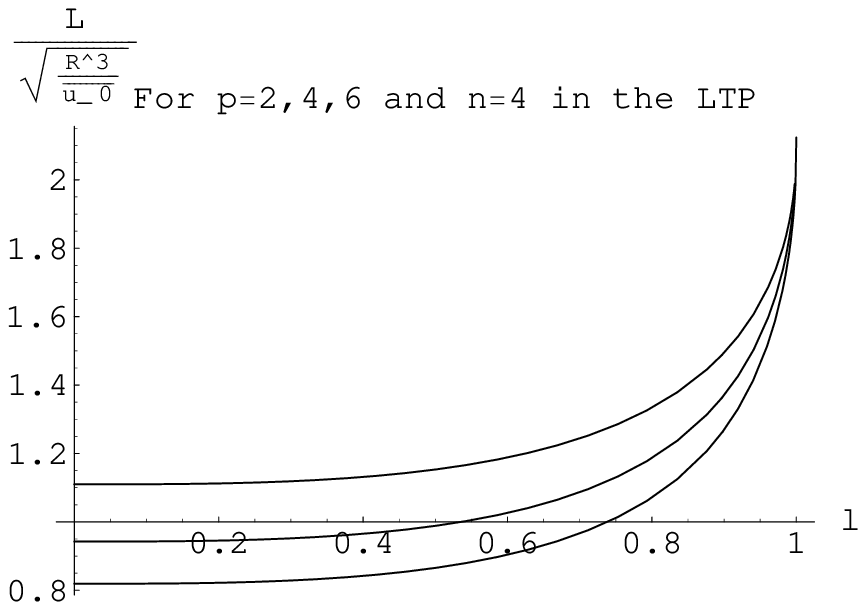}
\caption{$\f{L}{\sqrt{R^3/{\bar u}_0}}$ is plotted against $l=\f{u_0}{{\bar u}_0}$ for different flavor branes, D2-${\bar D2}$, D4-${\bar D4}$ and D6-${\bar D6}$ for the $QCD_4$ case an din the zero temperature phase.}
\label{fig_n_9_ltp}
\end{figure}

\begin{figure}[ht]
\includegraphics{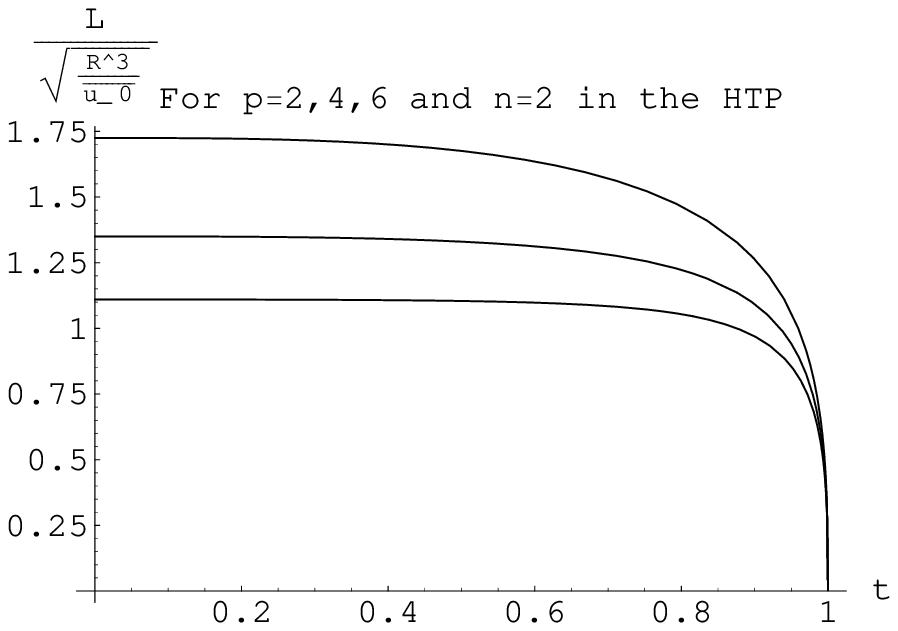}
\caption{$\f{L}{\sqrt{R^3/{\bar u}_0}}$ is plotted against $t=\f{u_T}{{\bar u}_0}$ for D2-${\bar D2}$, D4-${\bar D4}$ and D6-${\bar D6}$ in the $QCD_2$ case
and in the finite temperature phase.}
\label{fig_n_10_ltp}
\end{figure}
\begin{figure}[ht]
\includegraphics{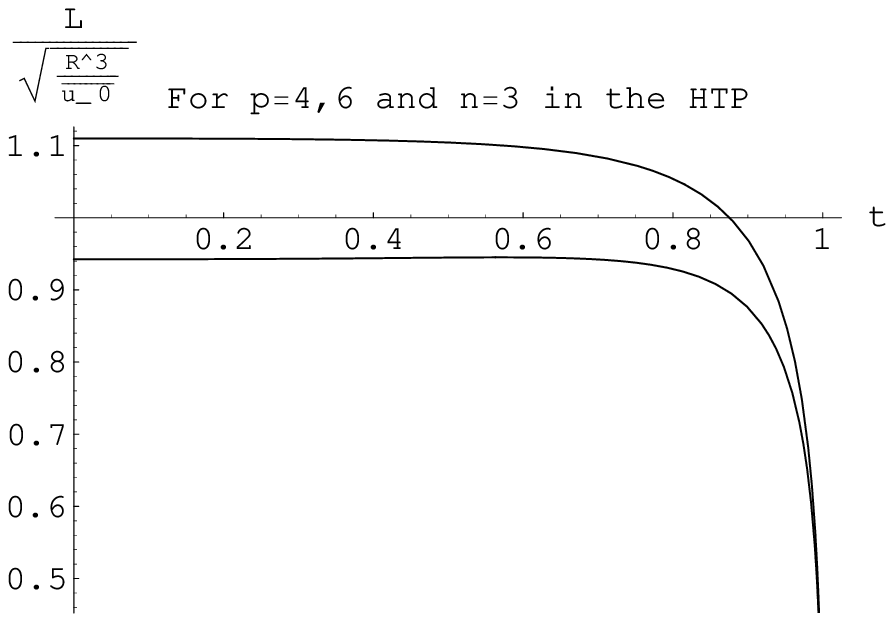}
\caption{$\f{L}{\sqrt{R^3/{\bar u}_0}}$ is plotted against $t=\f{u_T}{{\bar u}_0}$ for D4-${\bar D4}$ and D6-${\bar D6}$,  $QCD_3$ an din the finite temperature phase.}
\label{fig_n_11_ltp}
\end{figure}
\begin{figure}[ht]
\includegraphics{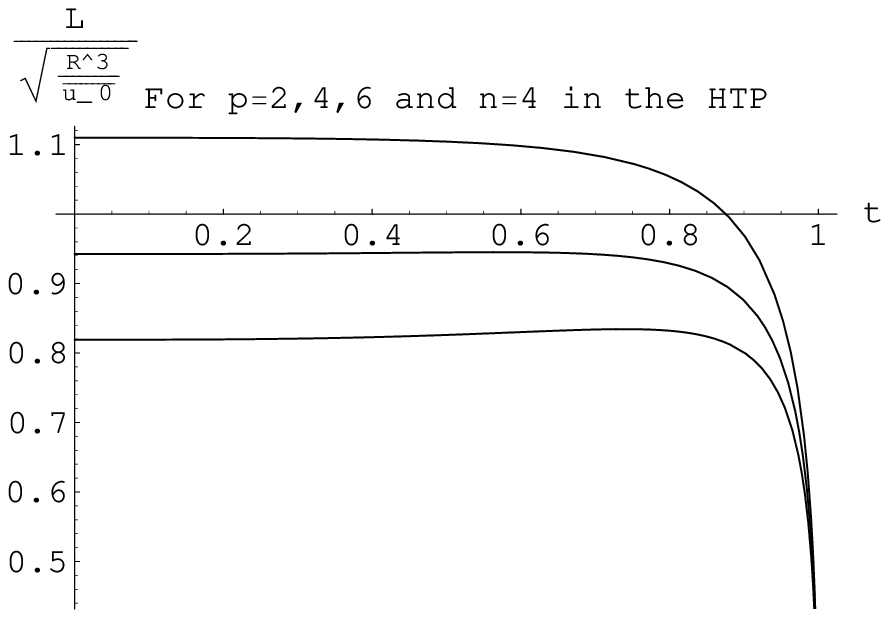}
\caption{$\f{L}{\sqrt{R^3/{\bar u}_0}}$ is plotted against $t=\f{u_T}{{\bar u}_0}$ for flavor branes: D2-${\bar D2}$, D4-${\bar D4}$ and D6-${\bar D6}$ for 
the $QCD_4$ case and in the finite temperature phase.}
\label{fig_n_12_ltp}
\end{figure}

Let us summarize the behavior of chiral symmetry breaking with the 
dependence on n and p and mention the 
value of $t$ where this happens, the temperature, the asymptotic separation 
between the flavor Dp-branes and anti-Dp branes, the temperature at the 
transition point and some dimensionless numbers. This is written down in 
eq \ref{brane_result_htp}.  \\

\begin{tabular}{|l|l|l|l|l|l|l|c|}
\hline
$QCD_n$&Dp-${\bar Dp}$&$t_c=(\f{u_T}{{\bar u}_0})_c$&$T_c \sqrt{\f{R^3}{{\bar u}_0}}$& $ \f{L_c}{\sqrt{R^3/{\bar u}_0}}$& $L_c T_c$& $\f{L_c}{R_{\tau}}$&\\
(n)&(p)&&&&&&\\
\hline
2&2&0.47951&0.165314&1.68167&0.278&1.74675&\\
\hline
2&4&0.55317&0.177558&1.3218&0.234696&1.47464&\\
\hline
2&6&0.613&0.186914&1.09676&0.205&1.28805&\\
\hline\hline
3&4&0.613&0.186914&1.09676&0.205&1.28805&\\
\hline
3&6&0.6622&0.19427&0.9439&0.183371&1.15216&\\
\hline\hline
4&4&0.6621&0.194255&0.943903&0.183358&1.15207&\\
\hline
4&6&0.7025&0.2&0.833931&0.166786&1.04795&\\
\hline
4&8&0.73573&0.204772&0.751283&0.153842&0.966616&\\
\hline\hline
\end{tabular}\\
\be
\label{brane_result_htp}
\ee

In general for  a given type of color brane,  in our case the temperature
is related to $ t=(\f{u_T}{{\bar u}_0})$ as
\be
T=\f{3}{4\pi} \sqrt{t}\f{1}{\sqrt{R^3/{\bar u}_0}}.
\ee
Denoting $T_c$ as the critical temperature at which there occurs phase 
transition and imposing the condition that at   this temperature there also
occurs a confinement-deconfinement phase transition. The confinement-deconfinement 
 gives an inverse  relation 
between the critical temperature $T_c$ and the radius of the compact circle 
$R_{\tau}$. 
\be
T_c=\f{1}{2\pi R_{\tau}}.
\ee

From figures (\ref{fig_n_7_ltp}), (\ref{fig_n_8_ltp}),  and 
(\ref{fig_n_9_ltp}) 
it can observed that the asymptotic separation
between the flavor branes increases as  the turning point ${\bar u}_0$ 
 approaches the minimum value i.e. $u_0$. which means the end points of the 
flavor Dp-branes and anti-Dp branes move away from each other as we allow 
these flavor branes to move deep into the bulk and when these flavor branes stay close to the boundary their separation takes the minimum value. 
This is true in the zero temperature phase and is true for all $n$ of 
$QCD_n$.  However, in the finite temperature case the situation is just 
opposite. As long as these flavor Dp-branes and anti-Dp branes stay close to 
the boundary their asymptotic separation attains maximum value and as these 
flavor branes move deep into the bulk and comes close to the horizon their
asymptotic separation between the end points decreases and takes a minimum
value at the horizon.

The absolute value of the difference between the actions 
$||S_{Dp}(U)-S_{Dp}(||)||$ for flavor branes in  the zero temperature case
 starts from a non-zero value then  increases 
attains a maximum value and then decreases towards zero. It means that deep
in the bulk both the actions take same value at $u_0$ for the zero temperature 
case, which it should by construction,  and at the horizon for the high temperature case.  But there is some thing
markedly different happens to this difference only at finite temperature phase 
that is for some special value of the turning point the actions take same 
value, even before reaching the horizon
and  the difference between the actions vanish. It is this point where 
there happens a phase transition. The  phase transition is first order.

From the table (\ref{brane_result_htp}), we see that this transition point 
depends on the number of common  directions that these flavor branes share 
with the color
branes   along with the dimensionality of the flavor brane. For a given n the
transition point increases with the increases of the dimensionality of the
flavor brane. From the table we also see that the transition point for a 
given n
with the maximum value of p almost coincides with the minimum 
value of p for n+1. The
transition temperature in a given unit increases with the increase of p and n.
Whereas the asymptotic distance of separation between the flavor Dp-brane
and anti-Dp brane decreases with the increase of p  for a given n. The
dimensionless quantity $L_c T_c$ decreases with the increase of p, so also
the ratio between the asymptotic separation of flavor branes to the radius
of the compact $\tau$ circle. 

\section{D4+D0 background}

Recently, there has been various attempts to study the phase diagram of various
geometric backgrounds by introducing charges through boosting the simple 
solutions  and also been used to generate black hole solutions 
with different charges. Here we shall adopt the same technique to generate 
additional charges to the D4 branes. The prescription is to boost the solution
of D4-branes along $t,\tau$ directions and then uplift the solution to 
11-dimension and reducing the solution along $\tau$ direction to generate the 
new solution which is electrically charged. The result will be same if we 
interchange the first two steps. The extra direction that we add while 
uplifting the solution to 11-dimension is taken to satisfy anti-periodic 
boundary conditions for fermions after reducing it to  10-dimension. 

The system that we are going to study is a non-confining background generated 
from the  background studied in the previous sections 
by introducing another charge, which would 
be D0-brane charge, by first boosting the solution and then uplifting it to
11-dimension and reducing it to generate a D4+D0 brane solution. We are
interested to study the chiral symmetry restoration in the above 
mentioned probe brane approximation with
temperature and with the D0-brane charge. To simplify our life henceforth,  
we shall be saying  the boost parameter not the D0-brane charge. 

The non-confining and near extremal solution is generated from the confining
solution by a double Wick rotation, $t\rightarrow -it, \tau\rightarrow -i\tau$
of (\ref{con_d4}). Now applying the above mentioned procedure yields
\beqn
\label{boosted_bh_solution}
& &ds^2=(M/h)^{1/2}(u)[-H dt^2+dx^2_1+dx^2_2+dx^2_3+dy^2]+(Mh)^{1/2}(u)[\f{du^2}{f}+u^2 d\Omega^2_4], \nn \\
& &e^{\phi}=(M^3/h)^{1/4}(u), \quad C_1=\f{(1-f(u))}{M(u)} sh_{\alpha_T} ch_{\alpha_T} 
dt,\quad H(u)=f/M \nn \\
& &  F_4=\frac{2\pi N_c}{V_4}\epsilon_4, \quad f(u)=1-(\frac{u_T}{u})^3, \quad
h=(R/u)^3, \quad M(u)= 1+(u_T/u)^3 (sh_{\alpha_T})^2.\nn \\
\eeqn
 
The boost parameter\footnote {Since $\tau$ is compact the boost should be 
thought to have taken in the the covering space. To avoid cluttering of the trigonometric function, we are using $Sinhx=sh_x, Coshx=ch_x$ etc.} $\alpha_T$
is as usual and  can take value from 0 to $\infty$. It is interesting to note 
that the periodicity of the Euclidean time circle is 
\be
\label{periodicity_of_boosted_t}
t\sim t+\beta, \quad \beta=\beta_o ch_{\alpha_T}, \quad \beta_o=\f{4\pi}{3} (\f{R^3}{u_T})^{(1/2)}
\ee
and the periodicity associated to the Euclidean time of the near extremal 
black hole solution generated by double Wick rotation of (\ref{con_d4})
is $\beta_0$, which is 
the zero boost limit of (\ref{periodicity_of_boosted_t}). 

Let us go through this procedure of boosting, uplifting and then reducing 
to generate an extra electric charge for the confining solution (\ref{con_d4}).
The result of all this generates a solution which is non-confining due to the
presence of electric charge. The explicit form of the solution is
\beqn
\label{boosted_nonbh_solution}
& &ds^2=(B/h)^{1/2}(u)[-F dt^2+dx^2_1+dx^2_2+dx^2_3+dy^2]+(Bh)^{1/2}(u)[\f{du^2}{f}+u^2 d\Omega^2_4], \nn \\
& &e^{\phi}=(B^3/h)^{1/4}(u), \quad C_1=\f{(f(u)-1)}{B(u)} sh_{\alpha_0} ch_{\alpha_0} 
dt,\quad F(u)=f/B \nn \\
& &  F_4=\frac{2\pi N_c}{V_4}\epsilon_4, \quad f(u)=1-(\frac{u_0}{u})^3, \quad
h=(R/u)^3, \quad B(u)= 1-(u_0/u)^3 (ch_{\alpha_0})^2.\nn \\
\eeqn

The radial coordinate u stays from $u_0$ to $\infty$ and when u take the lowest
value $u_0$, B becomes negative and the solution do not make much sense as the
dilaton becomes complex and the metric contains an overall phase. So, we shall be only dealing with eq \ref{boosted_bh_solution} for our study of chiral 
symmetry breaking. Also, while reducing along $\tau$ the periodicity of it
is not $2\pi$.

\subsection{Chiral symmetry restoration}

We shall  study chiral symmetry restoration  by putting a stack of
coincident D8-brane and anti-D8 brane on the compact direction $y$. As before 
there arises two different brane embeddings one which is straight and the 
other one which is curved. In the former case each set of coincident branes
do not do much whereas in the latter case there is a turning point for some 
value of radial coordinate $u={\bar u}_0$ and the end result is that there
is U-shaped configuration and it is this configuration of probe brane gives
the chiral symmetry breaking. Stating all these in a different way is: the 
gauge symmetry on the probe D8-branes is interpreted from the D4+D0-brane
point of view as a global symmetry which is nothing but the chiral symmetry.
For simplicity our calculation will  be done by taking a brane and 
anti-brane. 

The probe D8-brane action which we shall use has the as usual DBI action    
and there won't be any CS action as we are considering zero U(1) flux on the
worldvolume of the brane. It is easy to see that this is a consistent solution.
The induced metric on the D8 brane, in the static gauge choice and exciting
the lone scalar along the radial direction gives 
\be
ds^2_{D8}=(M/h)^{1/2}(u)[-H dt^2+dx^2_1+dx^2_2+dx^2_3]+(Mh)^{1/2}u^2 d\Omega^2_4+\f{(M/h)^{1/2}}{u^{'2}} (1+\f{h}{f} u^{'2}) du^2,
\ee
with $u^{'}=\f{du}{dy}$. The D8-brane action in the Euclidean signature
 becomes
\beqn
S_{D8}&=&T_8\int dt dx_1 dx_2 dx_3 du d^4\Omega e^{-\phi}\sqrt{-det g_8}\nn \\
&=&T_8 V_{3+1} V_4\int dy (HM^3)^{(1/2)} u^4 \sqrt{1+\f{h}{f}u^{'2}},
\eeqn
where $V_{3+1}, V_4$ are the volumes associated to $R^{1,3}, S^4$ respectively.
Note we have assumed $g_s=1$, for simplicity. The Lagrangian of the D8 brane
do not depend explicitly on $y$ coordinate, which means there is a conserved
quantity and is
\be
\f{d}{dy}(\f{{\sqrt{HM^3}u^4}}{\sqrt{1+\f{h}{f}u^{'2}}})=0.
\ee

The boundary condition at $y=0$ for the $U$-shaped solution is 
$u(0)={\bar u}_0$ and the condition on the 'velocity' is $u^{'}(0)=0$. With 
this condition the velocity 
\be
u^{'}=\sqrt{f/h}[\f{f}{f_0} (\f{M}{M_0})^2 (\f{u}{{\bar u}_0})^8-1]^{1/2},
\ee 
where $f_0=1-(\f{u_T}{{\bar u}_0})^3$ and $M_0=1+(\f{u_T}{{\bar u}_0})^3 
sh^2\alpha_T$. The value of u for which the D8 branes turns around and merge
 with the anti D8brane is  ${\bar u}_0$.

For the straight probe brane configuration in which the chiral symmetry is 
unbroken has got $f_0=0$, which means $M_0\rightarrow ch^2\alpha_T, 
H_0\rightarrow 0$. 
 
The energy associated to these configurations can be derived from the 
value of the action. Rewriting the action for the straight and U-shaped
configurations, with appropriate ranges for the integrals one ends up with   
\beqn
S_{D8}(||)&=&-[~.~]\int^{\infty}_{u_T} u^{(-1/2)} (u^3+u^3_T sh^2\alpha_T)\nn \\
S_{D8}(U)&=& -[~.~]\int^{\infty}_{{\bar u}_0} \f{(u^3+u^3_Tsh^2\alpha_T)^2 
u^{\f{-1}{2}} \sqrt{u^3-u^3_T}}{\sqrt{(u^3-u^3_T)(u^3+u^3_Tsh^2\alpha_T)^2-uf_0M^2_0{\bar u}^8_0}},
\eeqn
where $[~.~]=T_8V_{3+1}V_4 R^{3/2}$. By doing some change of variables 
and defining
$t=\f{u_T}{{\bar u}_0}$, we can integrate the U-shaped configuration and 
express the  integral in terms of a variable which lies 
 from 0 to 1, whereas for the straight
configuration the u integral can be written in terms of two pieces one 
from $u_T$ to ${\bar u}_0$ and the other is from ${\bar u}_0$ to $\infty$. The
result of doing all these gives us
\beqn
& &\Delta S_{D8}=S_{D8}(U)-S_{D8}(||)=[~.~]{\bar u^{7/2}}_0[-(2/7) (1-t^{7/2})-2sh^2\alpha_T(t^3-t^{7/2})+\f{1}{3}\int^1_0 dz\nn \\& & z^{-13/6}(1+t^3 z sh^2\alpha_T)\Bigg(\sqrt{\f{(1-t^3z)(1+t^3zsh^2\alpha_T)^2}{{(1-t^3 z)(1+t^3zsh^2\alpha_T)^2-(1-t^3)(1+t^3sh^2\alpha_T)^2z^{8/3}}}}-1\Bigg)]\nn \\
\eeqn  

Let the asymptotic distance between the D8-brane and anti D8-brane be 
denoted as L, which  depends on the boost parameter, and t as
\be
L=\f{2}{3}\sqrt{\f{R^3}{{\bar u}_0}}\sqrt{1-t^3} (1+t^3 sh^2\alpha_T)\int^1_0 
dz \f{z^{1/2}}{(1+t^3 z sh^2\alpha_T)\sqrt{1-t^3 z}\sqrt{(1-t^3z)-(1-t^3) z^{8/3} }}
\ee  

\begin{figure}[ht]
\includegraphics{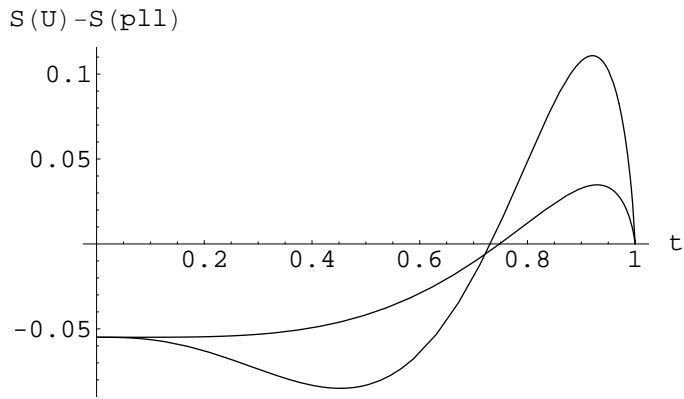}
\caption{The difference between the two allowed configuration, 
$S(U)-S(||)$, is plotted against $t=\f{u_T}{{\bar u}_0}$ for  D8-${\bar D8}$ 
flavor branes for 
the $QCD_4$ case and in the finite temperature phase for $\alpha_T=0.01,~1.0$.}
\label{fig_n_13_htp}
\end{figure}

\begin{figure}[ht]
\includegraphics{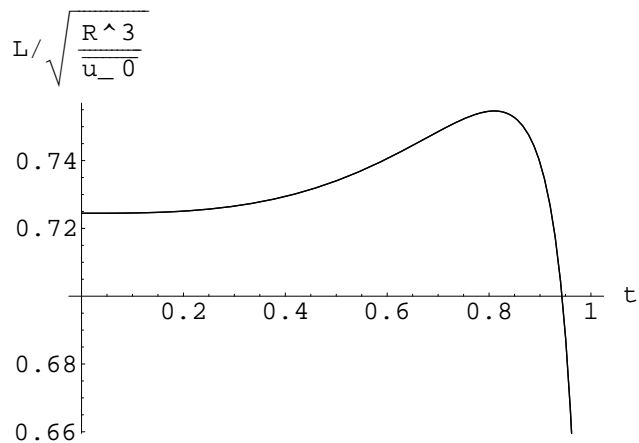}
\caption{$\f{L}{\sqrt{R^3/{\bar u}_0}}$ is plotted against $t=\f{u_T}{{\bar u}_0}$ for D8-${\bar D8}$ flavor branes  for 
the $QCD_4$ case and in the finite temperature phase for $\alpha_T=0.01,~1.0$.}
\label{fig_n_14_htp}
\end{figure}

The chiral symmetry restoration in this case can be seen from the figure (\ref{fig_n_13_htp}) for two different boosts, $\alpha_T=0.01, 1$, 

while that of the asymptotic separation, L,  between the D8-branes and 
anti-D8 
branes, when plotted is showing the usual behavior that we saw in 
earlier sections, is that when the flavor brane moves very deep into 
the bulk i.e. into the 
horizon the  L decreases and it increases when the flavor branes stay close 
to the boundary. This can be seen from the figure \ref{fig_n_14_htp}.
 
\section{Spacetime transition}

In this section we would like to study whether there would be any spacetime
transition between different geometries. There is indeed a transition between 
the confining geometry (\ref{con_d4}) and the finite temperature version of 
it, (\ref{n_x_bh}). However, the 
transition between either    (\ref{con_d4}) and (\ref{boosted_bh_solution})
or between  (\ref{n_x_bh}) and (\ref{boosted_bh_solution}) do not looks 
possible due to the presence of extra charge associated to boosting. But, 
that's not correct. We shall show by explicit computation that there indeed
happens a spacetime transition. 

The on-shell action for a set of coincident D4-branes which gives 
confinement, when evaluated in the Einstein frame with Euclidean time gives
\be
\label{action_con}
S_{confinement}=V_4 V_3 \beta 2\pi R_{\tau} \f{3}{16k^2} (\f{2\pi N_c}{V_4})^2
R^{-6} \int^{R_{\star}}_{u_0} du u^2,
\ee
where $V_4$ comes from integrating the four unit sphere, $S^4,~V_3$ comes from integrating the  three flat non-compact directions, $\beta$ is the periodicity of Euclidean time, $2\pi R_{\tau}$ is the periodicity associated to the 
compact $\tau$ direction. The rest of the terms in  (\ref{action_con}) comes 
from evaluating the Ricci curvature, the dilaton kinetic term and the 4-form 
term. The integral is being regulated as there is a UV divergence and is taken  from the minimum value of u which is $ u_0$ to some large value of u, $R_{\star}$.

Now evaluating the on-shell action for the boosted near extremal solution 
 (\ref{boosted_bh_solution}), in the   Einstein frame with Euclidean 
time is
\be
\label{boosted_bh_solution_action}
S_{Boosted}=V_4 V_3 \beta 2\pi R_y  \f{3}{16k^2}\int^{R_{\star}}_{u_T} du
[(\f{3u^2 u^3_T sh_{\alpha_T}ch_{\alpha_T}}{u^3+u^3_T sh^2_{\alpha_T}})^2 
u^{-2}+3 (\f{2\pi N_c}{V_4})^2 R^{-6} u^2 ].
\ee
Again the factors like $V_4, V_3, \beta$ arise from $S^4$, three flat 
directions and from the periodicity of Euclidean time direction respectively.
The periodicity of the compact $y$ direction is $2\pi R_y$. First term in 
(\ref{boosted_bh_solution_action}) arises, roughly, from the 2-from field. 
The range of the radial coordinate is from $u_T$ to a UV cut-off   $R_{\star}$.

Computing the periodicities following Wittens prescription \cite{ew}
we find that the 
divergences associated to each action are same and equal. So, by taking the 
difference between the actions one ends up with
\be
\label{diff_actions}
S_{Boosted}-S_{confinement}=V_4 V_3 \beta_0 ch_{\alpha_T} 2\pi R_{\tau}  
\f{3}{16k^2} \f{3u^3_T}{8\pi^3}[( u_0/u_T)^3-1+(\f{3}{4}+
8\pi^3) sh^2_{\alpha_T}]. 
\ee
 
It is easy to see that for for zero boost there is a phase transition for 
$ u_0=u_T$, which is the case between the confining and deconfining phase, 
\cite{asy}. It means when $u_0$ is smaller then $u_T$, the near extremal 
black hole without any D0-brane charge is preferred and when  $u_T$ becomes
smaller then $u_0$, the confined or zero temperature phase is preferred. 



It is interesting to note that the square bracket in (\ref{diff_actions}) can 
be written as 
\be
\label{constraint_boost}
(\f{\beta_0}{2\pi R_{\tau}})^6-1+(\f{3}{4}+8\pi^3)sh^2{\alpha_T}=
\left\{\begin{array}{ll}
-ve &\mbox{ for $\alpha_T < 0.064$ and for some value of $\f{\beta_0}{2\pi R_{\tau}}$}\\
+ve &\mbox{for $\alpha_T \geq 0.064$ and for any value of  $\f{\beta_0}{2\pi R_{\tau}}$}
\end{array}
\right.
\ee

The conclusion of this spacetime transition is that for the boost above 0.064 
and for any value to $\beta_0/2\pi R_{\tau}$, the preferred phase is the 
confining phase or the
zero temperature phase whereas if the boost is below 0.064 and $\beta_0/2\pi R_{\tau}$ is
less than 0.093 then the preferred phase is  that of the near extremal geometry
of D4+D0 brane. If the boost is below 0.064 but $\beta_0/2\pi R_{\tau}$ is 
above 0.093 then the preferred phase is the zero temperature phase. 

There appears a puzzle: for boost $\alpha_T=1.0$ we see from figure 
\ref{fig_n_13_htp}, 
there occurs a chiral symmetry restoration but from  
eq.(\ref{constraint_boost})
we see that for a boost above 0.064 the system stays in the confining phase.
Then how come we have chiral symmetry restoration in the confining phase? 

\section{Conclusion}

We have studied the Sakai-Sugimoto model of chiral symmetry breaking and 
restoration in the near horizon limit of $N_c$ color D4 brane using Dp-brane
and anti-Dp brane as flavor branes in the holographic setting. The result of 
phase transition, the asymptotic distance between the quarks $L$, depends
on the dimensionality of the flavor branes as well as on the number of 
directions  that color and flavor brane share. If the number of common 
directions that they share is $n$, then the quantities that is mentioned above
depends on p and n in a typical way. It appears that it goes as p+2n. So, for
p=6, n=2 and p=4, n=3, the phase diagrams and the distance L are identical.

The distance L, at zero temperature behaves in a different way than  at finite 
temperature. At zero temperature L increases as we move away from the boundary.
It takes maximum value deep in  the bulk. Whereas at finite temperature L 
decreases but very slowly as we move away from the boundary and it starts to
decrease very fast when we approach the horizon. The transition point, 
which describes chiral symmetry restoration, increases with the increase of
the dimensionality of the flavor brane in a given n, so also  the 
critical temperature  associated to this transition point.

The chiral symmetry restoration is described by a curve, which relates the 
asymptotic distance  of separation between the quarks with the 
temperature. In general this quantity is very difficult to compute but if
we evaluate it numerical then the curve is described by an equation $L T=c$,
where c is a constant and is much smaller than one. It means  for $L/R_{\tau}$ 
above $2\pi c$ there occurs the deconfined phase along with the chiral symmetry
restored phase. 

There appears two puzzles in this study of confinement-deconfinement 
transition 
and chiral symmetry breaking and restoration. One is the absence of NG boson
for the $QCD_2$ when p=2,6 and $QCD_3$ for D4-${\bar D4}$ case. This is due
to the absence of a normalisable zero mode $\phi_0$. There could be a 
possibility to avoid this for  $QCD_3$ case, as in odd dimensions one can't
have  Weyl fermions.    
The other is the simultaneous existence of
confinement and chiral symmetry restoration phase for boost above 0.064.

{\bf Note added}: As we were finishing the paper there appeared \cite{ahk}, 
with which there are some overlaps.  

{\bf Acknowledgments:}
SSP would like to  thank O. Aharony and especially J. Sonnenschein for 
useful discussions 
on the subject and also to the department of particle physics,   
Feinberg graduate school, WIS 
 for providing a supportive atmosphere.  Thanks to A. Babichenko for those 
physics and non-physics discussions. 
\section{Appendix}

\subsection{Chiral symmetry breaking}
In this section we shall study in detail the chiral symmetry breaking as 
discussed in the
introduction in the strong coupling limit. The near horizon bulk geometry 
(\ref{con_d4})of $N_c$ D4 branes is 
to be probed by a bunch of coincident Dp and anti-Dp branes, for p=4,6,8. The 
worldvolume direction of D4 are along $x^{\mu},u$, and that of D6 brane are
$x^{\mu}, u, $ and wrapped over the $S^2$ of $S^4$. The D8 brane is wrapped 
along $S^4$ and extended along $x^{\mu}, u$. The $S^4$ is described by 
 $d\Omega^2_4=d\theta^2_1+sin^2\theta_1 d\theta^2_2+
sin^2\theta_1 sin^2\theta_2 d\Omega^2_2$. D6 branes are kept at 
$\theta_2=\theta_1=\pi/2$. This means that there is no Chern-Simon part of
the action to probe Dp-brane. For single probe brane the action is now 
described by only the DBI part
\be
S_{Dp}=-\f{T_p V_4 V_{p-4}}{g_s} R^{3(p-8)/4}\int du u^{(p+8)/4} 
\sqrt{f(\f{d\tau}{du})^2+(\f{R}{u})^3\f{1}{f}}.
\ee 

We have excited  the only scalar $\tau$, radially. Its easy to see that the 
zero momentum configuration is a solution and it corresponds to a 
U-shaped configuration. It means the 
 branes and anti-branes gets joined   at the minimum value to u which is
$u_0$. By a slight abuse of notation we shall denote this by $||$.
There exists another configuration which is that of U-shaped but with 
non-zero momentum.
The boundary condition on $u(\tau)$: as 
$u(\tau=0)\rightarrow {\bar u}_0$ and $\f{du}{d\tau}|_{\tau=0}=0$, which makes
the U-shaped configuration  symmetric. 

The difference to the actions in the Minkowskian signature is
\beqn
S_{Dp}(U)-S_{Dp}(||)&=&-[]\bigg[\int^1_0 dz z^{-(p+18)/12}\bigg(\f{1}{\sqrt{1-l^3 z-(1-l^3) z^{(p+8)/6}}} -\f{1}{\sqrt{1-l^3 z}}\bigg)-\nn \\& &\int^{\f{1}{l^3}}_1  \f{z^{-(p+18)/12}}{\sqrt{1-l^3 z}}  \bigg],
\eeqn

where the U-shaped configuration with non-zero momentum has been integrated 
over the radial direction $u$ from the turning point 
${\bar u}_0$ to $\infty$ and the U-shaped with zero momentum  i.e. 
$||$-configuration is integrated from 
$u_0$ to $\infty$. Also, we have the changed the variable and 
$l=\f{u_0}{{\bar u}_0}$. The factor [~] comes from the prefactor of the Dp 
brane and some other numerical factor that appear from the change of variables and more importantly these terms are positive. 

In general its very difficult to find the difference analytically. So, we do
it numerically using Mathematica package. The difference is plotted in the
following figure(\ref{fig3}).

\begin{figure}[htb]
\includegraphics{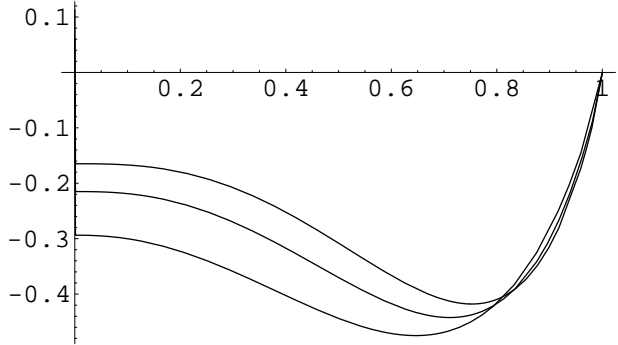}
\caption{$S_{Dp}(||)-S_{Dp}(U)$ is plotted against $l=\f{u_0}{{\bar u}_0}$ for p=4,6,8.}
\label{fig3}
\end{figure}
 
It is interesting to note that in Minkowskian signature $S_{Dp}(U)-S_{Dp}(||)
\sim  E(||)-E(U)$, for static configurations. Hence from the plot we 
conclude that for the embeddings in which the D4, D6 and D8 branes along
with their antibranes are stuck to the $\tau$ direction,   
prefers to take the 
U-shaped configuration, which is staying away from the $u_0$ line, there by 
 minimizing its energy. 
In the zero temperature case the 
 chiral symmetry $U(N_f)\times U(N_f)$ is broken to its diagonal subgroup 
$U(N_f)$ with the above choice of embeddings. 

\subsection{Goldstone boson}

The breakdown of the global symmetry implies there must appear Nambu-Goldstone
boson. Since, we start with the case with zero gauge potential, which is a 
solution to equation of motion. Now fluctuate the gauge potential with the 
assumption that we have only turned on the gauge potential along $A_{\mu}$ and
$A_u$. No gauge potential along the sphere directions. The quadratic 
fluctuation to gauge potential is governed by 
\be
\label{fl_A}
S_{Dp}=[]\int d^4x du[\f{K_1(u)}{4} \eta^{\mu\nu}\eta^{\rho\sigma}F_{\nu\rho}F_{\sigma\mu}+\f{K_2(u)}{2}\eta^{\mu\nu} F_{\nu u}F_{u\mu}],
\ee
where
\beqn
\f{K_1(u)}{4}&=&\f{u^{(p-4)/4} \sqrt{f}}{u^{'}} \sqrt{1+(\f{R}{u})^3\f{u^{'2}}{f^2}}\nn \\
\f{K_2(u)}{4}&=&\f{u^{(p-4)/4}u^{'}} {\sqrt{f}} \f{1}{\sqrt{1+(\f{R}{u})^3\f{u^{'2}}{f^2}}}.
\eeqn

The gauge potentials $A_{\mu}(x^{\mu},u)$ and $A_u(x^{\mu},u)$ are assumed to 
be expanded in terms  of complete sets of ortho-normal functions  as i.e. it is
assumed that this expansion solves the equation motion that follows from the
quadratic action (\ref{fl_A})
\beqn
& &A_{\mu}(x^{\mu},u)=\sum_{n=1} B^{(n)}_{\mu} (x^{\mu}) \psi_n(u),\quad A_{u}(x^{\mu},u)= \sum_{n=0} \pi^{(n)}_{\mu} (x^{\mu}) \phi_n(u)\nn \\& & 
F_{\mu\nu}=\sum_nF^{(n)}_{\mu\nu}(x^{\mu})\psi_n(u),\quad F_{\mu u}=\sum_n (\partial_{\mu}\pi^{(n)}(x)\phi_n(u)-B^{(n)}_{\mu}(x)\partial_u\psi_n(u),
\eeqn
where $F{(x)}_{\mu\nu}=\partial_{\mu}B^{(n)}_{\nu}-\mu\leftrightarrow\nu$. Let us also assume the following eigenvalue equation and the orthonormalisation
condition
\beqn
& &-\partial_u(K_2 \partial_u\psi_n)=K_1\lambda_n\psi_n,\quad [~]R^3\int du K_1 \psi_n\psi_m=\delta_{nm}\nn \\& &
[~]R^3\int du K_2\phi_m\phi_n=\delta_{mn} {\sf for ~all~ n}
\eeqn
The condition on $\phi_n$ gives a canonically normalized kinetic term. The 
condition hold  for $n \geq 1$, using $\lambda_n=m^2_n$
\be
\label{ident}
[~]R^3\int du K_2\partial_u\psi_n\partial_u\psi_m=\lambda_n\delta_{nm},\quad \partial_u\psi_n=\sqrt{\lambda_n} \phi_n=m_n\phi_n, \quad \phi_n=\sqrt{\f{K_1}{K_2}}\psi_n.
\ee

Substituting the above decomposition and the eigenvalue equation along the 
ortho-normalization condition and some identification from (\ref{ident}), one
ends up with

\be
S_{Dp}=-[~]R^3\int d^4x[\bigg(\sum_{n=1} \f{1}{4} F^{(n)}_{\mu\nu}F^{(n)}_{\rho\sigma}\eta^{\mu\rho}\eta^{\sigma\nu} +\f{1}{2}m^2_n B^{(n)}_{\mu} B^{(n)}_{\nu}\eta^{\mu\nu} \bigg)+\f{1}{2}\partial_{\mu}\pi^{(0)}\partial_{\nu}\pi^{(0)}\eta^{\mu\nu}], 
\ee
where we have made a gauge transformation associated to $ B^{(n)}_{\mu}$. 
From the 
above expression it follows that the massless gauge boson becomes massive by 
eating the $\pi^{(n)}$ for $n\geq 1$ and one left with the massless 
$\pi^{(0)}$,
which is interpreted as the Nambu-Goldstone boson associated to the chiral
symmetry breaking.

\subsection{Chiral symmetry restoration}
In this section we shall study the strong coupling analysis of the dual field 
theory to the background studied in the previous section in the context of 
probe brane approximation but in the high temperature or non-extremal solution 
of a bunch of coincident D4 branes. To begin, let us do a
Wick continuation of the solution
(\ref{con_d4}) and the final  non-extremal solution in Euclidean space is

 \beqn
\label{noncon_d4}
& &ds^2=(u/R)^{3/2}(f (u) dt^2+\sum^3_1\delta_{ij} dx^{i}dx^{j}+d\tau^2)+(R/u)^{3/2}(\frac{du^2}{f(u)}+u^2d\Omega^2_4),\nn \\
& &e^{\phi}=g_s (u/R)^{3/4},\quad F_4=\frac{2\pi N_c}{V_4}\epsilon_4, \quad f(u)=1-(\frac{u_T}{u})^3.
\eeqn

The periodicity of the Euclidean solution is 
\be
\beta=\f{4\pi}{3} (\f{R^3}{u_T})^{(1/2)}.
\ee

We want to probe this background using $D4-{\bar {D4}}, D6-{\bar {D6}}, 
D8-{\bar {D8}}$ branes i.e. a bunch of coincident  Dp and anti-Dp branes have 
been put at two different points along the $\tau$ circle for p=4, 6 and 8.
The DBI action that govern a part of the dynamics of these branes in this
background is
\be
\label{dbi_dp}
S_{DBI}=\f{T_p}{g_s}V_4 V_{p-4}R^{3(p-8)/4}\int du u^{(p+8)/4} {\sqrt f} 
\sqrt{(\f{d\tau}{du})^2+(R/u)^3 \f{1}{f}}.
\ee
   
It is interesting to note that we are considering a single Dp -anti Dp brane. 
The D4 branes are extended along $t, x^i, u$, D6 branes are extended along 
$t, x^i, u$ and wrapped along an $S^2$ of $S^4$. Keeping things a bit 
explicitly, the $d\Omega^2_4=d\theta^2_1+sin^2\theta_1 d\theta^2_2+
sin^2\theta_1 sin^2\theta_2 d\Omega^2_2$. D6 branes are kept at 
$\theta_2=\theta_1=
\pi/2$. The term $V_4$ that appear in the DBI action comes from integrating 
suitably $t, x^i$ directions, $V_{p-4}$ comes from integrating the appropriate part of $S^4$, sphere integral. Also, we have excited only one scalar field 
$\tau$ along the radial direction u.  There is no CS action for the case we 
are interested in. 

From (\ref{dbi_dp}) it follows trivially that
\be
\f{du}{d\tau}=[\f{u^{(p+8)/2} f}{{\bar u}^{(p+8)/2}_0 f_0}-1]^{(1/2)} {\sqrt f} 
(u/R)^{(3/2)},
\ee

where we have chosen the boundary condition as  $u(\tau=0)\rightarrow 
{\bar u}_0$ and $u^{\prime}(\tau=0)=0$. Here prime denotes derivative with respect to $\tau$ and $f_0=1-(\f{u_T}{{\bar u}_0})^3$.
The vanishing of the derivative of $u$ means there is a turning point and 
the configuration of the flavor brane looks: it turned around at ${\bar u}_0$
and joins with the
anti Dp branes forming a U-shaped configuration. There exists another 
configuration which is a zero 'momentum' configuration with $u^{'}$ blowing
up or the vanishing of $\f{d\tau}{du}$. The previous U-shaped  configuration 
from the point of view of $N_c$ D4 branes is that there is break down of 
$U(N_f)\times U(N_f)$ to its diagonal subgroup, whereas the latter zero 
momentum configuration implies there is no break down of this global symmetry.
The following conclusion can be drawn from these two kinds of configuration.
If the U-shaped configuration has got lower energy then there is a break down
of chiral 
symmetry and if the zero momentum configuration has got lower energy then 
there is no break down of chiral symmetry. 

Substituting back the expression of $u^{'}$ into the action and doing change 
of variables, one ends up with following actions for the U-shaped and zero
momentum configuration
\beqn
S(U-shaped)&=&[] (1/3)\int^1_0 dz z^{-(p+18)/12} (\f{1-t^3 z}{1-t^3 z-(1-t^3) z^{(p+8)/6}})^{(1/2)}\nn \\
S(||-shaped)&=&[]\bigg((4/p+6)[1-t^{(p+6)/4}]+ (1/3)\int^1_0 z^{-(p+18)/12}  \bigg),
\eeqn

Note, we have used the notation $||$-shaped to designate the zero momentum 
configuration, also the meaning of [] is that there are some less important positive term comes from both the change of variable and from the prefactor that multiplies the action in (\ref{dbi_dp}). In calculating the integrals the 
range of integration for the 
U-shaped configuration is, the radial coordinate stays from ${\bar u}_0$ to 
$\infty$, 
whereas for the $||$-shaped the radial coordinate stays from $u_T$ to 
$\infty$. The parameter $t=\f{u_T}{{\bar u}_0}$. 

Finally taking the difference between the actions we end up with
\be
S(U)-S(||)=[](1/3)\bigg[ dz z^{-\f{p+18}{12}} \bigg( \sqrt{\f{1-t^3 z}{1-t^3 z-(1-t^3) z^{\f{p+8}{6}}}}-1\bigg)-\f{12}{p+6} (1-t^{\f{p+6}{4}})\bigg]. 
\ee

Analytically its very difficult to proceed with the above integral. So to draw
any conclusion we shall use Mathematica package to find the difference as a
function of t. Note that this parameter stays from $0$ to $1$. The zero 
value to t means that the  configuration is close to the boundary of 
the bulk geometry and the latter means the  configuration is touching the
horizon of the bulk geometry.    

\begin{figure}[htb]
\includegraphics{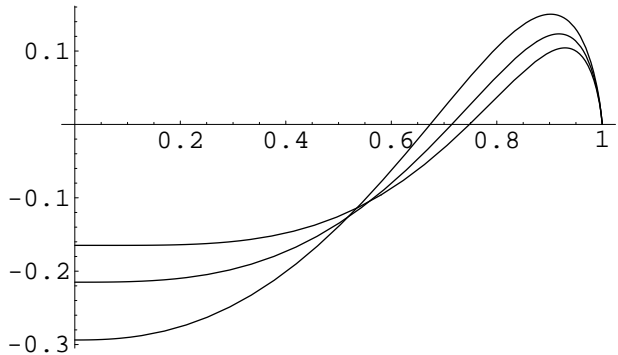}
\caption{$S_{Dp}(U)-S_{Dp}(||)$ is plotted against $t=\f{u_T}{{\bar u}_0}$ for p=4,6,8.}
\label{fig4}
\end{figure}

The difference between the actions for the U-shaped and parallel 
configuration is plotted in figure (\ref{fig4}).

For p=4, 6, 8, the critical value to $T$  are $t_c(p=4)=0.6221, t_c(p=6)=0.7025, t_c(p=8)=0.73573$. So, the pattern that emerges is that for increasing the the number of world-volume direction of the brane makes the critical value of 
$t_c$ to move towards right i.e. increases, which means the  configuration 
move towards the horizon of the bulk geometry. Naively, we can interpret 
this as: increasing the
value of p means increasing the tension of the brane and more massive it 
becomes and hence move away from the boundary.

Let the distance between the two Dp and anti-Dp  brane be L. Then this 
distance L is related to the turning point ${\bar u}_0$ and the horizon of the 
solution $u_T$ as 
\be
L=\f{2}{3} (R^3/{\bar u}_0)^{(1/2)} \sqrt{1-t^3} \int^1_0 dz \f{z^{\f{p-2}{12}}}{\sqrt{1-t^3 z}\sqrt{1-t^3 z-(1-t^3) z^{\f{p+8}{6}}}}.
\ee

Computing the distance of separation at the critical point $T_c$ we see that
the distance of separation decreases with the increase of the dimension of 
the world volume. For our case $L_c(p=4)=0.943903, L_c(p=6)=0.833931, L_c(p=8)=0.751283$ in units of $(R^3/{\bar u}_0)^{(1/2)}$. 

The critical inverse temperature at which there occurs  the chiral symmetry 
breaking is $\beta_c(p=4)5.14867, \beta_c(p=6)=5.0, \beta_c(p=8)=4.89065$
in units of $(R^3/{\bar u}_0)^{(1/2)}$. It means that the critical temperature at which there occurs chiral symmetry breaking increases with increasing the dimension
of the world-volume direction in units of $({\bar u}_0/R^3)^{(1/2)}$. 

The confinement-deconfinement transition occurs at temperature 
 $\beta_{c-dc}=2\pi R_{\tau}=1/T_{c-dc}$. Comparing both the confinement-deconfinement transition temperature and the chiral symmetry breaking temperature we find that $\f{L_c(p)}{R_{\tau}}$ decreases with increasing $p$, 
the dimension of the world-volume directions. For $\f{L_c(p=4)}{R_{\tau}}=1.15207, \f{L_c(p=6)}{R_{\tau}}=1.04795, \f{L_c(p=8)}{R_{\tau}}=0.966616$. From this analysis
we see that for the transition to happen the Dp brane and anti-Dp brane do 
not necessarily  sit at the diametrically opposite point of the $\tau$ 
circle.

The order of the transition from the chiral symmetry breaking phase to the 
chiral restoration phase is first order, which follows from the 
figure(\ref{fig4}). 
 
\section{Charging up the solution}

The charging up procedure that we follow is carried out by boosting and using
U-dualities. We start with a near extremal black hole solution in IIA  
and add an electric charge by first uplifting it to 11 dimension and then 
apply boost along the time and the $\tau$ direction. Finally reduce the 
solution along the $\tau$ direction to generate the solution which is 
charged under the electric field. 

Having spelled out the procedure let us move to find the form of the solution.
The starting point is the near extremal solution of a stack of coincident 
D4 brane 
\beqn
\label{n_x_bh}
& &ds^2_{10}=(u/R)^{3/2}(-f(u)dt^2+\sum^3_{i,j=1}\delta_{ij}dx^{i}dx^{j}+d\tau^2)+(R/u)^{3/2}(\frac{du^2}{f(u)}+u^2d\Omega^2_4),\nn \\
& &e^{\phi}=g_s (u/R)^{3/4},\quad F_4=\frac{2\pi N_c}{V_4}\epsilon_4, \quad f(u)=1-(\frac{u_T}{u})^3.
\eeqn  

Consider the Lorentz boost along $\tau$ direction as

\be
\left( \begin{array}{c} t_\textrm{new} \\ \tau_\textrm{new} \end{array} \right)
=
\left( \begin{array}{cc} \cosh\alpha_T & \sinh\alpha_T \\
                                   \sinh\alpha_T & \cosh\alpha_T \end{array} \right)
\left( \begin{array}{c} t_\textrm{old} \\ \tau_\textrm{old} \end{array} \right)
\ee

Henceforth, we shall drop the subscript from $(t,\tau)$. The metric in the 
11 dimensional space is
\be
\label{11d}
ds^2_{11}=e^{-2\phi/3}ds^2_{10}+e^{4\phi/3}(dy+A_{\mu}dx^{\mu})^2.
\ee
Using (\ref{n_x_bh}) in (\ref{11d}),  and boosting as mentioned above we find 
the 11 dimensional solution as 
\beqn
ds^2_{11}&=&e^{-2\phi/3}(R/u)^{-3/2}([-(L+\f{N^2}{M})dt^2+\sum^3_{i,j=1}\delta_{ij}dx^{i}dx^{j}]+M(d\tau+\f{N}{M}dt)^2)+\nn \\& &e^{-2\phi/3}(R/u)^{3/2}(\f{du^2}{f}+u^2 d\Omega^2_4)+e^{4\phi/3} dy^2,\nn \\
F_4&=&\frac{2\pi N_c}{V_4}\epsilon_4,
\eeqn
 where 
\be
M=1+(\f{u_T}{u})^3 sh^2\alpha_T,\quad N=(\f{u_T}{u})^3~ch\alpha_Tsh\alpha_T,\quad L=1-(\f{u_T}{u})^3 ch^2\alpha_T.
\ee
Reducing along $\tau$ direction with assuming $g_s=1$, gives 
\beqn
\label{boosted_bh_solution1}
& &ds^2=(M/h)^{1/2}(u)[-H dt^2+dx^2_1+dx^2_2+dx^2_3+dy^2]+(Mh)^{1/2}(u)[\f{du^2}{f}+u^2 d\Omega^2_4], \nn \\
& &e^{\phi}=(M^3/h)^{1/4}(u), \quad C_1=\f{(1-f(u))}{M(u)} sh_{\alpha_T} ch_{\alpha_T} 
dt,\quad H(u)=f/M \nn \\
& &  F_4=\frac{2\pi N_c}{V_4}\epsilon_4, \quad f(u)=1-(\frac{u_T}{u})^3, \quad
h=(R/u)^3, \quad M(u)= 1+(u_T/u)^3 (sh_{\alpha_T})^2.\nn \\
\eeqn

\end{document}